\documentclass[aps,prd,preprint,superscriptaddress,nofootinbib]{revtex4-2}
\usepackage{braket}
\usepackage{xargs}
\usepackage{mathtools}
\usepackage[english]{babel}
\usepackage[utf8]{inputenc}
\usepackage{amsmath,amssymb,braket,slashed,mathrsfs,amsthm}
\usepackage{pifont}
\usepackage{graphicx}
\usepackage{color,hyperref}
\usepackage{xcolor}
\usepackage{multirow,array}
\usepackage{bm}

\newcommand{\ba}{\begin{eqnarray}}
\newcommand{\ea}{\end{eqnarray}}

\newcommand\bnl{Physics Department, Brookhaven National Laboratory, Upton, NY 11973, USA}

\newcommand\riken{RIKEN-BNL Research Center, Brookhaven National Laboratory, Upton, NY 11973, USA}
\newcommand\edinb{School of Physics and Astronomy, The University of Edinburgh, Edinburgh EH9 3FD, UK}
\newcommand\uconn{Physics Department, University of Connecticut, Storrs, CT 06269-3046, USA}
\newcommand\soton{School of Physics and Astronomy, University of Southampton,  Southampton SO17 1BJ, UK}
\newcommand{\innovation}{Collaborative Innovation Center of Quantum Matter, Beijing 100871, China}
\newcommand{\chep}{Center for High Energy Physics, Peking University, Beijing 100871, China}
\newcommand{\pkuphy}{School of Physics, Peking University, Beijing 100871,
China}
\newcommand{\CERN}{CERN, Theoretical Physics Department, 1211 Geneva 23, Switzerland}
\newcommand{\liverpoolhopeU}{School of Mathematics, Computer Science and Engineering, Liverpool Hope University, Hope Park, Liverpool L16 9JD, UK}
\newcommand{\liverpoolU}{Theoretical Physics Division, Department of Mathematical Sciences, University of Liverpool, Liverpool L69 3BX, UK}

\begin{document}
\title{Lattice Calculation of Short-Range Contributions to Neutrinoless Double-Beta Decay 
$\pi^-\to\pi^+ ee$ at Physical Pion Mass}

\author{Peter~Boyle}\affiliation{\bnl}\affiliation{\edinb}

\author{Felix~Erben}\affiliation{\CERN}

\author{Xu~Feng}\affiliation{\pkuphy}\affiliation{\innovation}\affiliation{\chep}

\author{Jonathan~M.~Flynn}\affiliation{\soton}

\author{Nicolas~Garron}\affiliation{\liverpoolhopeU}\affiliation{\liverpoolU}

\author{Taku~Izubuchi}\affiliation{\bnl}\affiliation{\riken}

\author{Luchang~Jin}\affiliation{\uconn}

\author{Rajnandini~Mukherjee}\affiliation{\edinb}\affiliation{\soton}

\author{J.~Tobias~Tsang}\affiliation{\CERN}\affiliation{\liverpoolU}

\author{Xin-Yu~Tuo}\email{ttxxyy.tuo@gmail.com}\affiliation{\bnl}

\date{\today}

\preprint{CERN-TH-2025-132}

\begin{abstract}
Neutrinoless double-beta (\(0\nu\beta\beta\)) decays provide an excellent probe for determining whether neutrinos are Dirac or Majorana fermions.
The short-range matrix elements associated with the \(\pi^- \rightarrow \pi^+ ee\) process contribute at leading order in the \(0\nu\beta\beta\) decay channel \(nn \to ppee\) through pion exchange between nucleons.
However, current lattice calculations show notable discrepancies in predicting these short-range contributions.
To address this issue, we perform a lattice QCD calculation of the \(\pi^- \rightarrow \pi^+ ee\) matrix elements using domain wall fermion ensembles at the physical pion mass generated by the RBC/UKQCD Collaboration. To mitigate contamination from around‑the‑world effects, we develop a new method to reconstruct and subtract them directly from lattice data.
We then perform a nonperturbative renormalization using the RI/SMOM approach in \((\gamma_\mu,\gamma_\mu)\) and \((\slashed{q},\slashed{q})\) schemes.
Compared with previous studies, this work reduces the uncertainties in the matrix elements and provides an independent cross-check that helps to reconcile the discrepancies among previous lattice calculations.
\end{abstract}

\maketitle

\section{Introduction\label{sec1}}
Neutrino oscillation experiments have confirmed that neutrinos have nonzero masses \cite{Super-Kamiokande:1998kpq,SNO:2002tuh,KamLAND:2002uet,Strumia:2006db,RevModPhys.75.345}.
This phenomenon, which goes beyond the Standard Model, has made the nature of neutrino masses a significant focus in particle physics research.
A primary question is whether neutrinos are Dirac or Majorana fermions~\cite{Majorana:1937vz}.
Neutrinoless double-beta ($0\nu\beta\beta$) decay experiments \cite{GERDA:2020xhi,Majorana:2022udl,LEGEND:2021bnm,KamLAND-Zen:2022tow,CUORE:2021mvw,CUPID-0:2022yws,CUPID:2020aow,EXO-200:2019rkq} can distinguish between these two scenarios:
observation of $0\nu\beta\beta$ decay would establish that neutrinos have a Majorana mass term and provide information on the neutrino mass scale $m_{\beta\beta}$.
Moreover, $0\nu\beta\beta$ decay is a lepton-number-violating (LNV) process with $\Delta L=2$. Because lepton number violation can lead to baryon number violation \cite{PhysRevD.45.455,Davidson:2008bu}, it can also help explain the matter-antimatter asymmetry of the universe.

According to the underlying LNV mechanism, $0\nu\beta\beta$ decay has long-range contributions mediated by light Majorana neutrino exchange and short-range contributions arising from other LNV mechanisms at higher energy scales.
In the Standard Model Effective Field Theory (SMEFT), the former stems from the dimension-five Weinberg operator~\cite{Weinberg:1979sa}, which generates the Majorana mass term, while the latter arises from higher-dimensional effective operators~\cite{Babu:2001ex,Prezeau:2003xn,deGouvea:2007qla,Lehman:2014jma,Graesser:2016bpz,Cirigliano:2018yza}. These short-range contributions can also generate a Majorana mass term for neutrinos~\cite{Schechter:1981bd}.
The relative size of the short-range and long-range contributions depends on specific BSM scenarios.
For instance, if the energy scale $\Lambda$ of new physics can be as low as $O(\text{TeV})$ (e.g., from the mass $m_{\nu_R}$ of right-handed sterile neutrinos), then the short-range contributions are much less suppressed than for $\Lambda\gg 1~\text{TeV}$, whereas the long-range contribution is substantially suppressed by the neutrino mass scale $m_{\beta\beta}$, making both contributions potentially comparable in $0\nu\beta\beta$ decay.
Thus, to probe LNV mechanisms in BSM scenarios via $0\nu\beta\beta$ decay, it is crucial to accurately compute the matrix elements of these short-range contributions. 

In this work, we focus on the short-range contribution to $\pi^- \to \pi^+ ee$, which contributes to the $0\nu \beta\beta$ decay channel $nn \to ppee$ through pion exchange between nucleons. This contribution appears at leading order in the $\chi$EFT power counting~\cite{Prezeau:2003xn}. 
In contrast, the long-range contribution to $\pi^- \to \pi^+ ee$ mediated by neutrino exchange enters at next-to-next-to-leading order (NNLO) in $\chi$EFT~\cite{Cirigliano:2017tvr}.
(In addition to pion-exchange mechanisms, four-nucleon contact interactions also appear at leading order through renormalization~\cite{Cirigliano:2018hja}.) For the short-range contribution, the heavy degrees of freedom can be integrated out to obtain dimension-9 low-energy effective operators respecting the $SU(3)_c \times U(1)_{em}$ gauge symmetries, whose hadronic parts are described by local four-quark operators~\cite{Prezeau:2003xn,Graesser:2016bpz}:
\begin{equation}\label{Operators}
	\begin{aligned}
		& \mathcal{O}_{1+}^{++}=\left(\bar{q}_L  \tau^{+} \gamma^\mu q_L \right)\left[\bar{q}_R  \tau^{+} \gamma_\mu q_R \right] \\
		& \mathcal{O}_{2+}^{++} =\left(\bar{q}_R  \tau^{+} q_L \right)\left[\bar{q}_R  \tau^{+} q_L \right]+\left(\bar{q}_L  \tau^{+} q_R \right)\left[\bar{q}_L  \tau^{+} q_R \right] \\
		& \mathcal{O}_{3+}^{++} =\left(\bar{q}_L  \tau^{+} \gamma^\mu q_L \right)\left[\bar{q}_L  \tau^{+} \gamma_\mu q_L \right]+\left(\bar{q}_R  \tau^{+} \gamma^\mu q_R \right)\left[\bar{q}_R  \tau^{+} \gamma_\mu q_R \right] \\
		& \mathcal{O}_{1+}^{\prime ++} =\left(\bar{q}_L  \tau^{+} \gamma^\mu q_L \right]\left[\bar{q}_R  \tau^{+} \gamma_\mu q_R \right) \\
		& \mathcal{O}_{2+}^{\prime++} =\left(\bar{q}_R  \tau^{+} q_L \right]\left[\bar{q}_R  \tau^{+} q_L \right)+\left(\bar{q}_L  \tau^{+} q_R \right]\left[\bar{q}_L  \tau^{+} q_R \right),
	\end{aligned}
\end{equation}
where parentheses $(\cdots)$ and brackets $[\cdots]$ indicate color contractions for the quarks. $\tau^+ = \left( \begin{smallmatrix} 0 & 1 \\ 0 & 0 \end{smallmatrix} \right)$ is the isospin raising operator.
Among these operators, $\mathcal{O}_{1+}^{(\prime)++}$ and $\mathcal{O}_{2+}^{(\prime)++}$ are related to the LO contributions in $\chi$EFT, while $\mathcal{O}_{3+}^{++}$ appears at NNLO.
We omit operators with odd parity, which do not contribute to $\pi^- \to \pi^+ ee$, as well as vector operators that are suppressed by the small electron mass.

To obtain reliable inputs for EFT calculations, the hadronic matrix elements relevant to $0\nu 2\beta$ decays can be determined nonperturbatively using lattice QCD simulations. 
For the process $\pi^- \to \pi^+ ee$, previous lattice QCD calculations have investigated both the long-range contribution~\cite{Detmold:2020jqv,Feng:2018pdq,Tuo:2019bue} and the short-range contribution~\cite{Nicholson:2018mwc,Detmold:2022jwu}. 
However, the two lattice studies in Refs.~\cite{Nicholson:2018mwc,Detmold:2022jwu} reported significantly different results for the short-range contribution.
For instance, the bag parameter $B^{\overline{\mathrm{MS}}}_\pi(\mu)$ in the $\overline{\mathrm{MS}}$ scheme with scale $\mu$ is related to the operator $\mathcal{O}_3$ by $B^{\overline{\mathrm{MS}}}_\pi(\mu) = 2\langle\pi^+|\mathcal{O}_3(\mu)|\pi^-\rangle^{\overline{\mathrm{MS}}} / \bigl(\frac{8}{3} m_\pi^2 f_\pi^2\bigr)$, where $f_\pi$ is normalized according to the PDG convention with $f_\pi\approx 130$ MeV~\cite{ParticleDataGroup:2024cfk}.
At $\mu=3~$GeV, Ref.~\cite{Nicholson:2018mwc} quotes $B^{\overline{\mathrm{MS}}}_\pi(\mu)=0.421(23)$, whereas converting the matrix element $\langle\pi^+|\mathcal{O}_3(\mu)|\pi^-\rangle^{\overline{\mathrm{MS}}}$ reported in Ref.~\cite{Detmold:2022jwu} to the same definition gives $B^{\overline{\mathrm{MS}}}_\pi(\mu)=0.197(18)$. These two results disagree by approximately a factor of two. Therefore, an independent lattice calculation is needed to cross-check and clarify this discrepancy, ensuring consistent lattice QCD predictions for this contribution.

Motivated by this, we perform a lattice calculation of the matrix elements $\langle \pi^+|\mathcal{O}_i(\mu)|\pi^-\rangle^{\overline{\mathrm{MS}}}$ $(i=\{1,2,3,1^\prime,2^\prime\})$ in the $\overline{\mathrm{MS}}$ scheme with $\mu=3~$GeV at the physical pion mass.
We find that the backward propagation of the light pion introduces substantial around-the-world effects in the temporal direction, which spoil the plateau in the ratio of three-point to two-point correlation functions.
To resolve this, we propose a new subtraction method to directly reconstruct and remove these effects from the lattice data, thereby restoring a stable plateau.
%
%
The lattice matrix elements are renormalized nonperturbatively using the Rome-Southampton method with non-exceptional kinematics (RI/SMOM)~\cite{Sturm:2009kb}. The conversion to $\overline{\mathrm{MS}}$ is perturbative, with perturbative truncation effects estimated by applying two different RI/SMOM schemes, denoted by $(\gamma_\mu,\gamma_\mu)$ and $(\slashed q,\slashed q)$ (see Section~\ref{subsec:renormalization} and Refs.~\cite{Boyle:2024gge,Boyle:2017skn}).
The same renormalization coefficients were also used in the study of BSM kaon mixing~\cite{Boyle:2024gge}.

The remainder of this paper is organized as follows:
Section~\ref{sec2} provides a detailed description of the lattice methodology, including the subtraction of the around-the-world effect and the RI/SMOM renormalization.
In Section~\ref{sec3}, we present our numerical results and compare them with the existing literature~\cite{Nicholson:2018mwc,Detmold:2022jwu}.
Finally, we offer concluding remarks in Section~\ref{sec4}.
Further details on the renormalization coefficients used in our calculation, as well as additional discussions of the around-the-world effect, are given in the appendix.

\section{Methodology\label{sec2}}
\subsection{Bare Matrix Elements and Subtraction of Around-the-World Effects}
To compute the bare matrix elements (denoted as $\langle \mathcal{O}_i \rangle=\langle \pi^+|\mathcal{O}_i|\pi^-\rangle$), we evaluate the lattice two-point and three-point correlation functions:
\begin{equation}
	\begin{aligned}
		C_2(t) &= \langle \phi_\pi(t) \,\phi^{\dagger}_\pi(0)\rangle,\\
		C_{3}^i(t_1, t_2) &= \langle \phi_\pi^{\dagger}(t_2) \,\mathcal{O}_i(0)\, \phi^{\dagger}_\pi(-t_1)\rangle.
	\end{aligned}
\end{equation}
Here, the pion interpolating operator is defined as $\phi_\pi = \bar{u}\gamma_5 d$, and the time coordinates $t_1, t_2 \in [0, T)$ denote the temporal locations of the interpolating fields. To illustrate the necessity of addressing around-the-world effects in three-point functions, we first review two commonly used methods for extracting matrix elements from the ratio of three-point to two-point functions:
\begin{itemize}
	\item \textbf{Ratio Method 1 (denoted as \(R_1\))}: This method eliminates around-the-world effects in the two-point function and is defined by
	\begin{equation}
		O^{(R_1)}_i(t)
		= \frac{C_3^i(t,t)}{N_\pi^2 \,e^{-2m_\pi t}},
	\end{equation}
	where \(N_\pi = \vert \langle 0 \vert \phi_\pi \vert \pi \rangle \vert /(2 m_\pi)\) and \(m_\pi\) are extracted from the two-point function $C_2(t)=2m_\pi N_\pi^2 (e^{-m_\pi t}+e^{-m_\pi(T-t)})$. An equivalent formulation, employed in Ref.~\cite{Detmold:2022jwu}, is given by
	\begin{equation}
		O^{(R_1)}_i(t)
		= 2m_\pi\,\frac{C_3^i(t,t)}{C_2(2t)
		-\tfrac12\,C_2 (T/2)\,e^{m_\pi(2t - T/2)}},
	\end{equation}
	which directly cancels the around-the-world effects in \(C_2(2t)\) by incorporating the data at \(t = T/2\).

	\item \textbf{Ratio Method 2 (denoted as \(R_2\))}: This method is defined by
	\begin{equation}
		O^{(R_2)}_i(t)
		= (2m_\pi N_\pi)^2 \,
		\frac{C_3^i(t,t)}{C_2(t)\,C_2(T - t)}.
	\end{equation}
	It was employed in Ref.~\cite{Nicholson:2018mwc} to extract the $\pi^-\to\pi^+ ee$ matrix elements.
\end{itemize}

\begin{figure}
	\centering
	\includegraphics[width=0.65\textwidth]{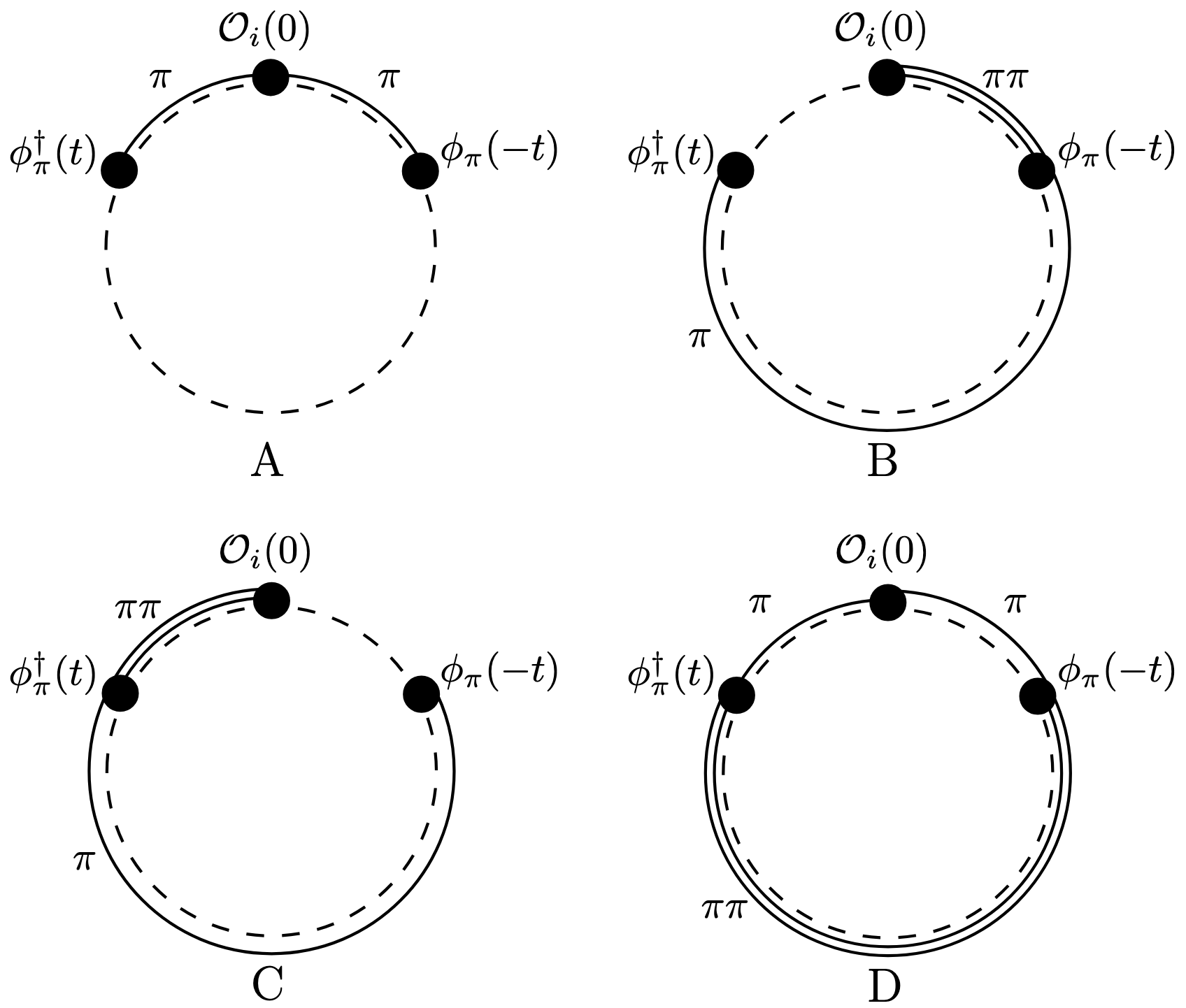}
	\caption{Illustration of around-the-world effects. Dashed lines indicate the periodic temporal direction, oriented counter-clockwise, while solid lines represent intermediate $\pi$ or $\pi\pi$ states. Diagram A is the contribution we aim to compute. In diagrams B and C, a single pion propagates across the temporal boundary, thereby generating around-the-world effects. In diagram D, two pions propagate across the temporal boundary, similarly producing around-the-world effects.
\label{fig:ATW}}
\end{figure}

Next, we analyze the around-the-world effects in these methods. Fig.~\ref{fig:ATW} shows various diagrams of pion backward propagation. In these diagrams, dashed lines represent the temporal direction under the periodic boundary condition, with the forward time direction oriented counter-clockwise. The solid lines denote $\pi$ or $\pi\pi$ intermediate states. 

Taking into account the around-the-world effects from diagrams B, C, and D, the quantity \(O^{(R_1)}_i(t)\) can be expressed as
\begin{equation}\label{OR1}
	\begin{aligned}
		O^{(R_1)}_i(t)
		&= N_\pi^{-2}e^{2 m_\pi t}\,\bigl(C^i_{3,A}(t, t)
		+ C^i_{3,B}(t, t)
		+ C^i_{3,C}(t, t)
		+ C^i_{3,D}(t, t)\bigr)\\
		&= \mathcal{N}_{A,i}
		+ (\mathcal{N}_{B,i} + \mathcal{N}_{C,i})
		\,e^{-m_\pi(T - 2t) - \Delta E_{\pi \pi}\,t}
		+ \mathcal{N}_{D,i}\,e^{-(2m_\pi + \Delta E_{\pi\pi})\,(T - 2t)}.
	\end{aligned}
\end{equation}
In this equation and throughout the following discussion, we consider only the dominant ground-state contribution and possible around-the-world effects from $\pi$ and $\pi\pi$ states. All higher excited-state contributions are neglected. $\mathcal{N}_{A,i}=\langle \mathcal{O}_i \rangle=\langle \pi \vert \mathcal{O}_i \vert \pi\rangle $ represents the bare matrix element to be computed. Time-reversal symmetry implies that the coefficients corresponding to diagrams B and C satisfy $\mathcal{N}_{B,i} = \mathcal{N}_{C,i} = \dfrac{\langle \pi \vert \phi_\pi^\dagger \vert \pi\pi\rangle}{\langle \pi \vert \phi_\pi^\dagger \vert 0\rangle}\,\langle \pi\pi \vert \mathcal{O}_i \vert 0\rangle$. The coefficient associated with diagram D is given by 
$\mathcal{N}_{D,i} 
= \dfrac{\bigl\lvert \langle \pi \vert \phi_\pi^\dagger \vert \pi\pi\rangle \bigr\rvert^2}{\bigl\lvert \langle \pi \vert \phi_\pi^\dagger \vert 0\rangle \bigr\rvert^2}
\,\langle \pi \vert \mathcal{O}_i \vert \pi\rangle $. We neglect normalization differences between single-particle and two-particle states. $\Delta E_{\pi\pi} = E_{\pi\pi} - 2m_\pi$ denotes the difference between the \(\pi\pi\) ground-state energy and $2m_\pi$ on the lattice. For the physical pion mass ensembles used in this work, we find that $\Delta E_{\pi\pi}/(2m_\pi) \sim 0.5\%$.

\begin{figure} 
	\centering
	\includegraphics[height=0.605\textwidth]{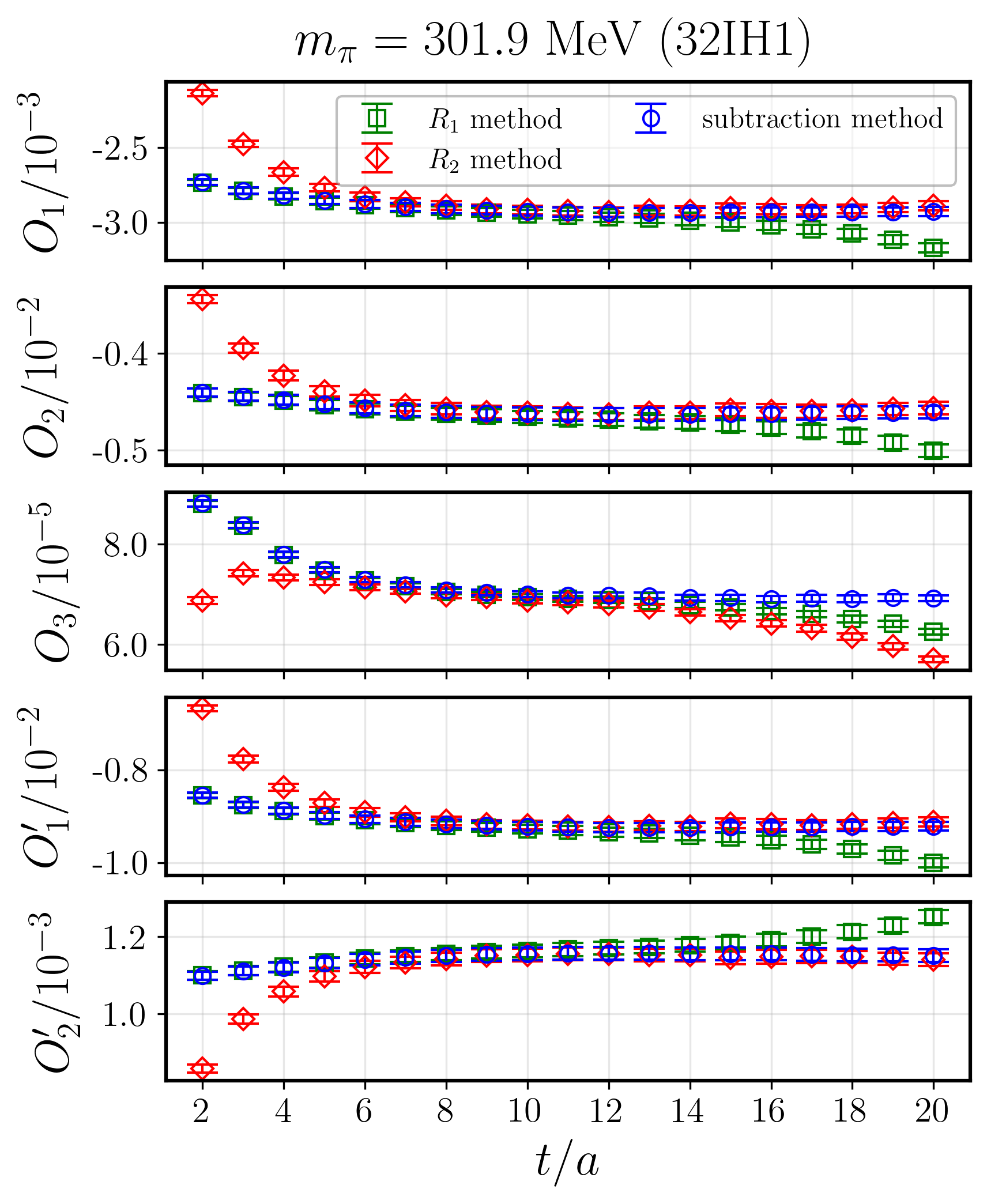}
	\includegraphics[height=0.605\textwidth]{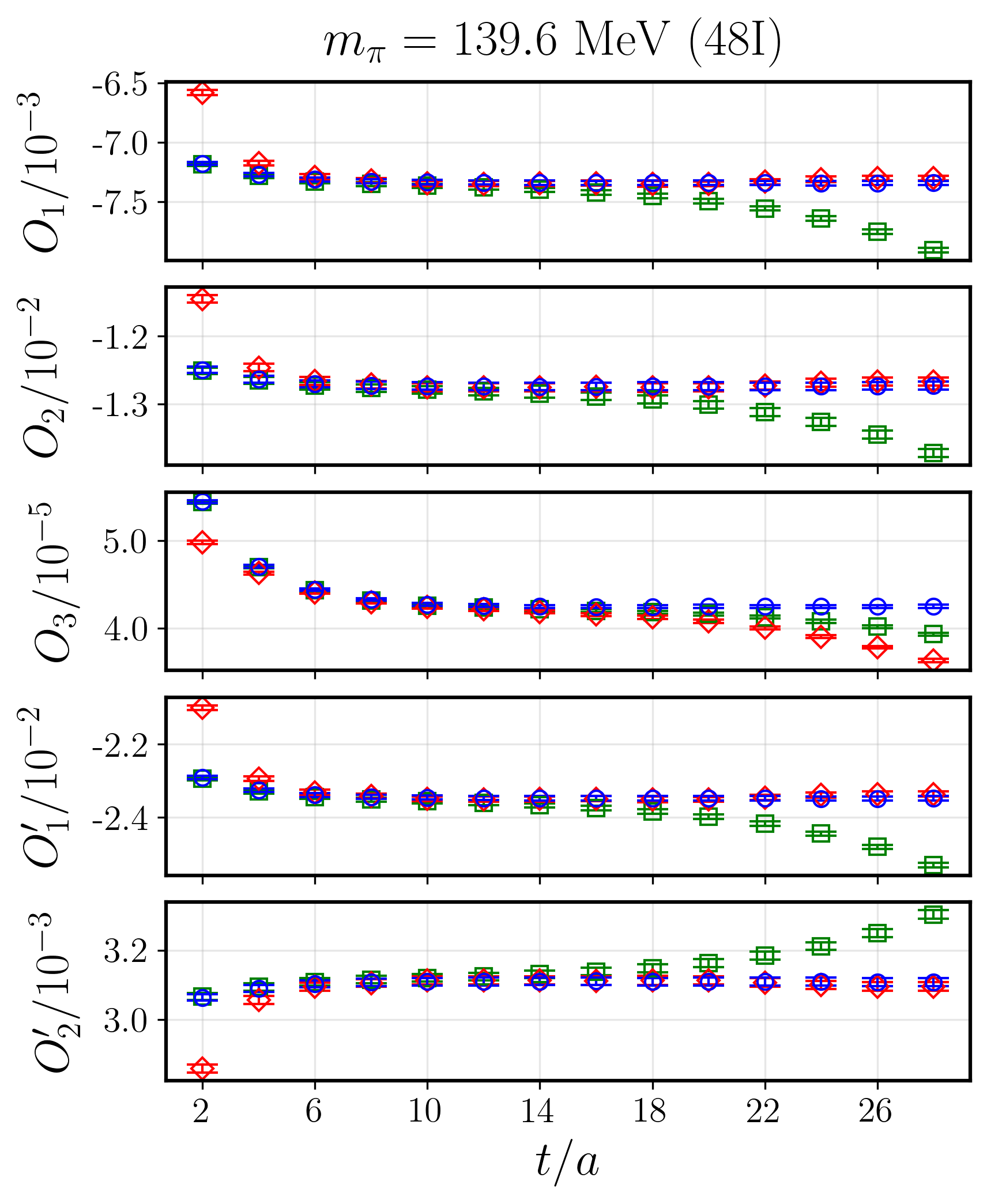}
	\caption{Comparison between the effective bare matrix elements before subtraction of around-the-world effects, \(O_i^{(R_1)}(t)\) (green squares) and \(O_i^{(R_2)}(t)\) (red diamonds), and after subtraction, \(O_i^{(\text{sub})}(t_{\mathrm{sep}})\) (blue circles). The left and right panels present results at unphysical and physical pion mass, respectively.
		\label{fig:methods}
	}
\end{figure}

From this expression, it follows that as \(t \to T/2\), the around-the-world effects arising from diagrams B, C, and D can introduce errors of \(\mathcal{O}(100\%)\). In Fig.~\ref{fig:methods}, we demonstrate that the around-the-world effects in the \(R_1\) method remain significant even for \(t \ll T/2\), as exemplified by the 32IH1 and 48I ensembles (see Table~\ref{table:ens} for ensemble parameters). These effects are sizable both in physical and unphysical pion mass ensembles. A more detailed analysis of the contributions of the individual diagrams in Fig.~\ref{fig:ATW} is presented in Appendix.~\ref{Append:ATW}.

Similarly, we examine the around-the-world effects in $O^{(R_2)}_i(t)$ for the $R_2$ method:
\begin{equation}
	O^{(R_2)}_i(t)
	= \frac{
		\mathcal{N}_{A,i}
		+ (\mathcal{N}_{B,i}+\mathcal{N}_{C,i}) 
		\,e^{-m_\pi(T-2t)
		-\Delta E_{\pi\pi}\,t}
		+ \mathcal{N}_{D,i}\,
		e^{-(2m_\pi+\Delta E_{\pi\pi})
		(T-2t)} 
	}{
		1 + 2\,e^{-m_\pi(T-2t)}
		+ e^{-2m_\pi(T-2t)}
	}.
\end{equation}
Under the approximations \(\Delta E_{\pi\pi} \approx 0\) and
\(\mathcal{N}_{A,i} \approx \mathcal{N}_{B,i} \approx \mathcal{N}_{C,i} \approx \mathcal{N}_{D,i}\), the around-the-world effects in the two-point and three-point functions cancel. However, the validity of this approximation depends on the specific operator. In Appendix~\ref{Append:ATW}, using actual lattice data, we find that for the matrix elements with \(i \in \{1,2,1',2'\}\), this approximation holds well. By contrast, for \(i = 3\), a more suitable approximation is 
\(\mathcal{N}_{A,3} 
\approx -\,\mathcal{N}_{B,3} 
\approx -\,\mathcal{N}_{C,3} 
\approx \mathcal{N}_{D,3}\). Consequently, as illustrated in Fig.~\ref{fig:methods}, the $R_2$ method achieves better suppression of around-the-world effects for the matrix elements with \(i \in \{1,2,1',2'\}\). In contrast, for the matrix element with \(i = 3\), the around-the-world effects are not suppressed. Therefore, the efficacy of the $R_2$ method in suppressing around-the-world effects depends on the specific matrix element.

\begin{figure}
	\centering
	\includegraphics[width=0.5\textwidth]{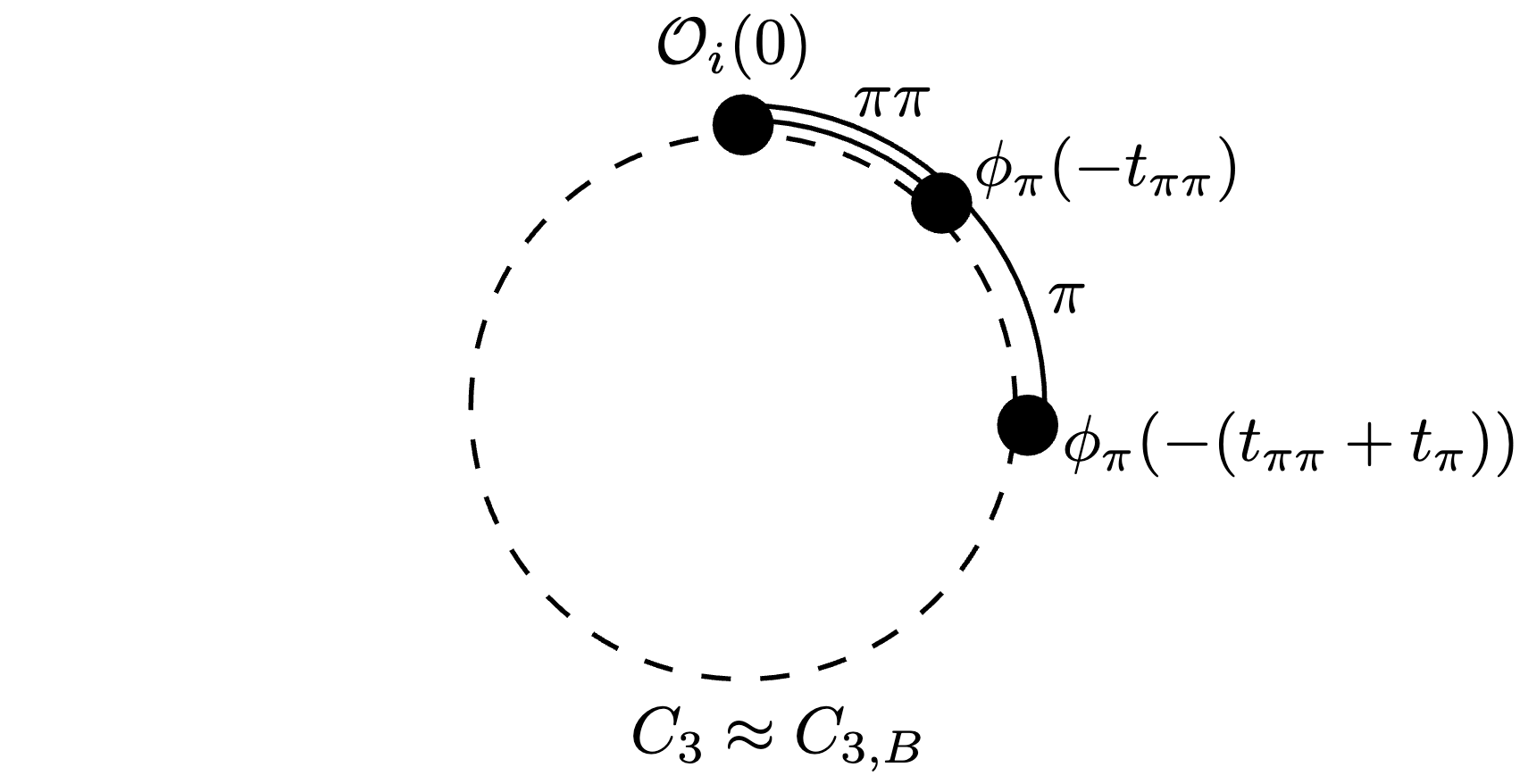}
	\caption{Extraction of the contribution from diagram~B under the time setup \((t_1, t_2) = (t_{\pi \pi}, T - t_{\pi\pi} - t_{\pi})\), for which the contribution from diagram~B dominates those from diagrams A, C, and D.\label{fig:Bsub}}
\end{figure}

In this work, we propose a new method to subtract the around-the-world effects from diagrams B and C. As illustrated in Fig.~\ref{fig:Bsub}, when \(0 \ll t_1 < (T - t_2) \ll T/2\), diagram B dominates over diagrams A, C, and D. Thus, for these time locations, we can directly extract the information from diagram B. To do this, we choose \((t_1, t_2) = (t_{\pi\pi}, T - t_{\pi\pi} - t_{\pi})\), where \(t_{\pi}\) and \(t_{\pi\pi}\) are large enough to ensure that the pion and the two-pion ground state dominate, respectively. When \(t_\pi + t_{\pi\pi} \ll T/2\), the three-point function is dominated by diagram~B:
\begin{equation}\label{Osub0}
	C^i_3\bigl(t_{\pi\pi}, T - t_{\pi\pi} - t_\pi\bigr)
	\approx C^i_{3,B}\bigl(t_{\pi\pi}, T - t_{\pi\pi} - t_\pi\bigr)
	= N_\pi^2\,\mathcal{N}_{B,i}\,e^{-m_\pi t_\pi - E_{\pi\pi}\,t_{\pi\pi}}.
\end{equation}

We can then define a quantity in which the around-the-world effects from diagrams~B and C are reconstructed from the data and then subtracted:
\begin{equation}\label{Osub}
	O_i^{(\text{sub})}(t)
	= N_\pi^{-2}e^{2 m_\pi t}\Bigl[C^i_3(t, t)
	- 2\,C^i_3\bigl(t_{\pi\pi},T - t_{\pi\pi} - t_\pi\bigr)\,
	e^{-m_\pi(T - t_\pi - 2t)}\,
	e^{-E_{\pi \pi}\,(t - t_{\pi\pi})}\Bigr].
\end{equation}

The validity of neglecting the (A, C, D) diagrams in 
$C^i_3\bigl(t_{\pi\pi},\,T-t_{\pi\pi}-t_\pi\bigr)$ 
can be examined using the 48I ensemble as an example. 
We take $t_\pi = t_{\pi\pi} = \Delta T = 1.4~\mathrm{fm}$ 
(corresponding to $a\Delta T = 12$) and estimate the ratios of the (A, C, D) diagrams to the dominant B diagram as
\begin{equation}
\begin{aligned}
    \frac{C^i_{3,A}(\Delta T, T-2\Delta T)}{C^i_{3,B}(\Delta T, T-2\Delta T)}
    &\approx \frac{e^{-m_\pi (T-\Delta T)}}{e^{-m_\pi \Delta T - E_{\pi\pi}\Delta T}}
    \approx 2\%,\\[3pt]
    \frac{C^i_{3,C}(\Delta T, T-2\Delta T)}{C^i_{3,B}(\Delta T, T-2\Delta T)}
    &\approx \frac{e^{-m_\pi \Delta T - E_{\pi\pi}(T-2\Delta T)}}{e^{-m_\pi \Delta T - E_{\pi\pi}\Delta T}}
    \approx 0.006\%,\\[3pt]
    \frac{C^i_{3,D}(\Delta T, T-2\Delta T)}{C^i_{3,B}(\Delta T, T-2\Delta T)}
    &\approx \frac{e^{-m_\pi (T-\Delta T)-E_{\pi\pi}\Delta T}}{e^{-m_\pi \Delta T - E_{\pi\pi}\Delta T}}
    \approx 0.3\%.
\end{aligned}
\end{equation}
In the plateau region $t \in [1.8,\,3.0]~\mathrm{fm}$ used for the 48I ensemble (see Fig.~\ref{fig:results3}), the around-the-world effects remain $\lesssim 5\%$. 
Consequently, omitting this $O(2\%)$ contribution from the $(A,C,D)$ diagrams induces a systematic uncertainty $\lesssim 0.1\%$, well below the statistical uncertainties of $O(0.3$–$0.5\%)$.

Moreover, since \(\Delta E_{\pi\pi}/E_{\pi\pi}\sim 0.5\%\), we can safely neglect this difference in correcting the around-the-world effects. By setting \(E_{\pi\pi} = 2m_\pi\), Eq.~\eqref{Osub} reduces to:
\begin{equation}\label{Osub2}
    O_i^{(\text{sub})}(t)
	= N_\pi^{-2}e^{2 m_\pi t}\Bigl[C^i_3\bigl(t,t\bigr)
	- 2\,C^i_3\bigl(t_{\pi\pi},T - t_{\pi\pi} - t_\pi\bigr)\,
	e^{-m_\pi(T - t_\pi - 2 t_{\pi\pi})}\Bigr].
\end{equation}

Table~\ref{table:sub} shows a comparison between setting \(\Delta E_{\pi\pi}\) to zero (i.e., \(E_{\pi\pi}=2m_\pi\)) and using the exact value \(\Delta E_{\pi\pi}=1.445(55)\,\mathrm{MeV}\) extracted from the two-point function \(C_{\pi\pi}(t) = \langle \phi^\dagger_{\pi\pi,I=2}(t)\,\phi_{\pi\pi,I=2}(0)\rangle\) in ensemble 48I. The results indicate that the impact of treating $\Delta E_{\pi\pi}$ as zero lies well below the level of statistical uncertainty and that this effect is therefore negligible. Eq.~(\ref{Osub2}) provides a convenient method for subtracting around-the-world effects, relying only on the three-point function and the pion mass as inputs.
\begin{table} 
	\centering
	\begin{tabular}{c|ccccc}
		\hline\hline
		\(\Delta E_{\pi\pi}\)
		& \(a^4\langle \mathcal{O}_1\rangle/10^{-3}\)
		& \(a^4\langle \mathcal{O}_2\rangle/10^{-2}\)
		& \(a^4\langle \mathcal{O}_3\rangle/10^{-5}\)
		& \(a^4\langle \mathcal{O}^\prime_{1}\rangle/10^{-2}\)
		& \(a^4\langle \mathcal{O}^\prime_{2}\rangle/10^{-3}\)\\
		\hline
		\(\Delta E_{\pi\pi}=0\)
		& \(-7.343(19)\)
		& \(-1.2741(57)\)
		& \(4.248(15)\)
		& \(-2.3481(59)\)
		& \(3.137(14)\)\\
		exact \(\Delta E_{\pi\pi}\)
		& \(-7.344(19)\)
		& \(-1.2744(57)\)
		& \(4.247(15)\)
		& \(-2.3487(59)\)
		& \(3.137(14)\)\\
		\hline\hline
	\end{tabular}
	\caption{\label{table:sub}
	Comparison of using \(\Delta E_{\pi\pi} = 0\) (Eq.~\eqref{Osub2}) vs.\ exact \(\Delta E_{\pi\pi} = 1.445(55)\,\mathrm{MeV}\) (Eq.~\eqref{Osub}) when subtracting around-the-world effects in ensemble 48I. The results indicate that treating \(\Delta E_{\pi\pi}\) as zero has an impact well below the current level of statistical uncertainty.}
\end{table}

In Fig.~\ref{fig:methods}, we compare the subtraction method \(O_i^{(\text{sub})}(t)\) (blue circles), with ratio methods \(O_i^{(R_1)}(t)\) (green squares) and \(O_i^{(R_2)}(t)\) (red diamonds). The left and right panels present results at unphysical and physical pion mass, respectively. From this comparison, we see that subtracting the contributions of diagrams~B and C greatly reduces the around-the-world effects, leading to better plateaux. This subtraction method works well because the around-the-world effects are dominated by diagrams~B and C for \(t \ll T/2\), which is demonstrated numerically in Appendix.~\ref{Append:ATW}.

\subsection{Renormalization}
\label{subsec:renormalization}

To obtain well-defined physical quantities in the continuum limit, the bare matrix elements computed on the lattice must be renormalized. In this work, we adopt the Rome-Southampton method with non-exceptional kinematics (RI/SMOM) for renormalization~\cite{Sturm:2009kb}. To assess the systematic uncertainty associated with the perturbative truncation of the matching from $\mathrm{RI}$ to $\overline{\mathrm{MS}}$, we apply two different RI/SMOM schemes: \((\gamma^\mu,\gamma^\mu)\) and \((\slashed{q},\slashed{q})\). We use the NPR basis for the four-quark operators:
\begin{equation}
	\begin{aligned}
		Q_1 &=(\bar{q}\,\tau^+ \gamma_\mu\,q)\,(\bar{q}\,\tau^+ \gamma^\mu\,q)
		+ (\bar{q}\,\tau^+\gamma_5\gamma_\mu\,q)\,(\bar{q}\,\tau^+\gamma_5\gamma^\mu\,q),\\
		Q_2 &=(\bar{q}\,\tau^+ \gamma_\mu\,q)\,(\bar{q}\,\tau^+\gamma^\mu\,q)
		- (\bar{q}\,\tau^+\gamma_5\gamma_\mu\,q)\,(\bar{q}\,\tau^+\gamma_5\gamma^\mu\,q),\\
		Q_3 &=(\bar{q}\,\tau^+ q)\,(\bar{q}\,\tau^+q)
		- (\bar{q}\,\tau^+\gamma_5\,q)\,(\bar{q}\,\tau^+\gamma_5\,q),\\
		Q_4 &=(\bar{q}\,\tau^+q)\,(\bar{q}\,\tau^+q)
		+ (\bar{q}\,\tau^+\gamma_5\,q)\,(\bar{q}\,\tau^+\gamma_5\,q),\\
		Q_5 &=\sum_{\mu<\nu}
		(\bar{q}\,\tau^+\gamma_\mu\gamma_\nu\,q)\,(\bar{q}\,\tau^+\gamma^\mu\gamma^\nu\,q).
	\end{aligned}
\end{equation}
The relationship between this basis and the BSM basis in Eq.~(\ref{Operators}) is~\cite{Detmold:2022jwu}:
\begin{equation}\label{QO}
	\left(\begin{array}{l}
		Q_1(x) \\
		Q_2(x) \\
		Q_3(x) \\
		Q_4(x) \\
		Q_5(x)
	\end{array}\right)=\left(\begin{array}{ccccc}
		0 & 0 & 2 & 0 & 0 \\
		4 & 0 & 0 & 0 & 0 \\
		0 & 0 & 0 & -2 & 0 \\
		0 & 2 & 0 & 0 & 0 \\
		0 & 2 & 0 & 0 & 4
	\end{array}\right)\left(\begin{array}{l}
		\mathcal{O}_1(x) \\
		\mathcal{O}_2(x) \\
		\mathcal{O}_3(x) \\
		\mathcal{O}^\prime_{1}(x) \\
		\mathcal{O}^\prime_{2}(x)
	\end{array}\right)
\end{equation}

In the RI/SMOM method, the renormalized matrix elements in the \(\overline{\mathrm{MS}}\) scheme are given by
\begin{equation}\label{NPR}
	\langle Q_n\rangle^{\overline{\mathrm{MS}}}(\mu,a)
	= R_{ni}^{\overline{\mathrm{MS}}\leftarrow \mathrm{RI}}(\mu)\,
	Z_{ij}^{\mathrm{RI}}(\mu,a)\,\langle Q_j\rangle(a),
\end{equation}
where \(\langle \cdots\rangle\) and \(\langle\cdots\rangle^{\overline{\mathrm{MS}}}\) represent the bare and renormalized matrix elements, respectively. The factor \(R_{ni}^{\overline{\mathrm{MS}}\leftarrow \mathrm{RI}}(\mu)\) is the one-loop matching coefficient from RI scheme to \(\overline{\mathrm{MS}}\) scheme, for which we use results from Ref.~\cite{Boyle:2017skn}. The matrix \(Z_{ij}^{\mathrm{RI}}(\mu,a)\) is the renormalization factor in the RI/SMOM scheme, which is given by
\begin{equation}
	\frac{Z_{ij}^{\mathrm{RI},(\mathcal{A}, \mathcal{B})}(\mu, a)}{Z_A^2(a)}
	\times
	\left.
	\lim_{m_q \rightarrow 0}
	\frac{P_k^{(\mathcal{A})}\Bigl[\Pi_j^{\text {bare }}(a, p_1, p_2)\Bigr]}
	{\Bigl(P_A^{(\mathcal{B})}\bigl[\Pi_A^{\text {bare }}(a, p_1, p_2)\bigr]\Bigr)^2}
	\right|_{\mathrm{SMOM}}
	= \frac{F_{ik}^{(\mathcal{A})}}{\bigl(F_A^{(\mathcal{B})}\bigr)^2},
\end{equation}
where \(\mathcal{A}\) and \(\mathcal{B}\) are either \(\gamma_\mu\) or \(\slashed{q}\), indicating the choice of RI/SMOM scheme. We compute the renormalization factors for \((\mathcal{A}, \mathcal{B})=(\gamma_\mu,\gamma_\mu)\) and \((\slashed{q},\slashed{q})\). The quantities \(\Pi_j^{\text {bare}}(a, p_1, p_2)\) represent the amputated vertex functions of the four-quark operators.
The projectors \(P_k^{(\mathcal{A}/\mathcal{B})}\) are specific to each scheme, while \(F_{ik}^{(\mathcal{A})}\) and \(F_A^{(\mathcal{B})}\) are the tree-level projected values.
%
For further details on the computation of \(Z_{ij}^{\mathrm{RI}}(\mu,a)\), including chiral extrapolation and step-scaling studies, see Refs.~\cite{Boyle:2017skn,Boyle:2024gge}. Numerical results for \(Z_{ij}^{\mathrm{RI}}(\mu=3~\mathrm{GeV},a)\) are presented in Appendix~\ref{Appendix:ZRI}. Note that in Ref.~\cite{Boyle:2024gge}, the renormalization coefficients \(Z_{ij}^{\mathrm{RI}}(\mu,a)\) are quoted in the SUSY basis, while we present them in the NPR basis.

\section{Numerical Results\label{sec3}}
\subsection{Lattice Setup}
We use $N_f=2+1$ domain wall fermion ensembles generated by the RBC/UKQCD collaboration \cite{RBC:2010qam,RBC:2014ntl}. Table \ref{table:ens} summarizes the parameters of these ensembles. The ensembles with physical pion masses (48I and 64I) have similar volumes but different lattice spacings, thereby enabling a direct continuum extrapolation to physical results. For comparison with the results in Ref.~\cite{Detmold:2022jwu}, we also compute the matrix elements in the same unphysical pion mass ensembles. In computing the correlation functions, we use Coulomb-gauge-fixed wall source propagators. We insert the operator $\mathcal{O}_i$ at every time slice and average over all the slices, using time-translation invariance for $N_{\text{conf}}\to\infty$ limit.

\begin{table}
	\centering
	\begin{tabular}{cccccc}
		\hline\hline
		Ensembles&$a^{-1}$[GeV]&$(L/a)^3\times (T/a)$&$am_l$&$m_\pi$[MeV]&$N_{\text{conf}}$\\
		\hline
		24IH1&1.7848(50)&$24^3\times 64$&0.005&341.0(8)&77\\
		24IH2&1.7848(50)&$24^3\times 64$&0.01&431.1(7)&76\\
		32IH1&2.3833(86)&$32^3\times 64$&0.004&301.9(1.1)&49\\
		32IH2&2.3833(86)&$32^3\times 64$&0.006&359.9(1.1)&49\\
		$48 \mathrm{I}$&1.7295(38)&$48^3\times 96$&0.00078&139.6(2)&110\\
		$64 \mathrm{I}$&2.3586(70)&$64^3\times 128$&0.000678&139.2(3)&33\\
		\hline\hline
	\end{tabular}
	\caption{\label{table:ens}Parameters of the lattice ensembles used in this study. For each ensemble, we provide the inverse lattice spacing $a^{-1}$ (in GeV), the lattice volume $(L/a)^3\times (T/a)$, the light quark mass $am_l$, the pion mass $m_\pi$, and the number of configurations $N_{\text{conf}}$ used in this work. The first four ensembles correspond to unphysical pion masses \cite{RBC:2010qam}, whereas the last two ensembles correspond to physical pion masses \cite{RBC:2014ntl}.
}
\end{table}

\subsection{Numerical Results for Bare Matrix Elements}
\begin{figure} 
	\centering
	\includegraphics[height=0.55\textwidth]{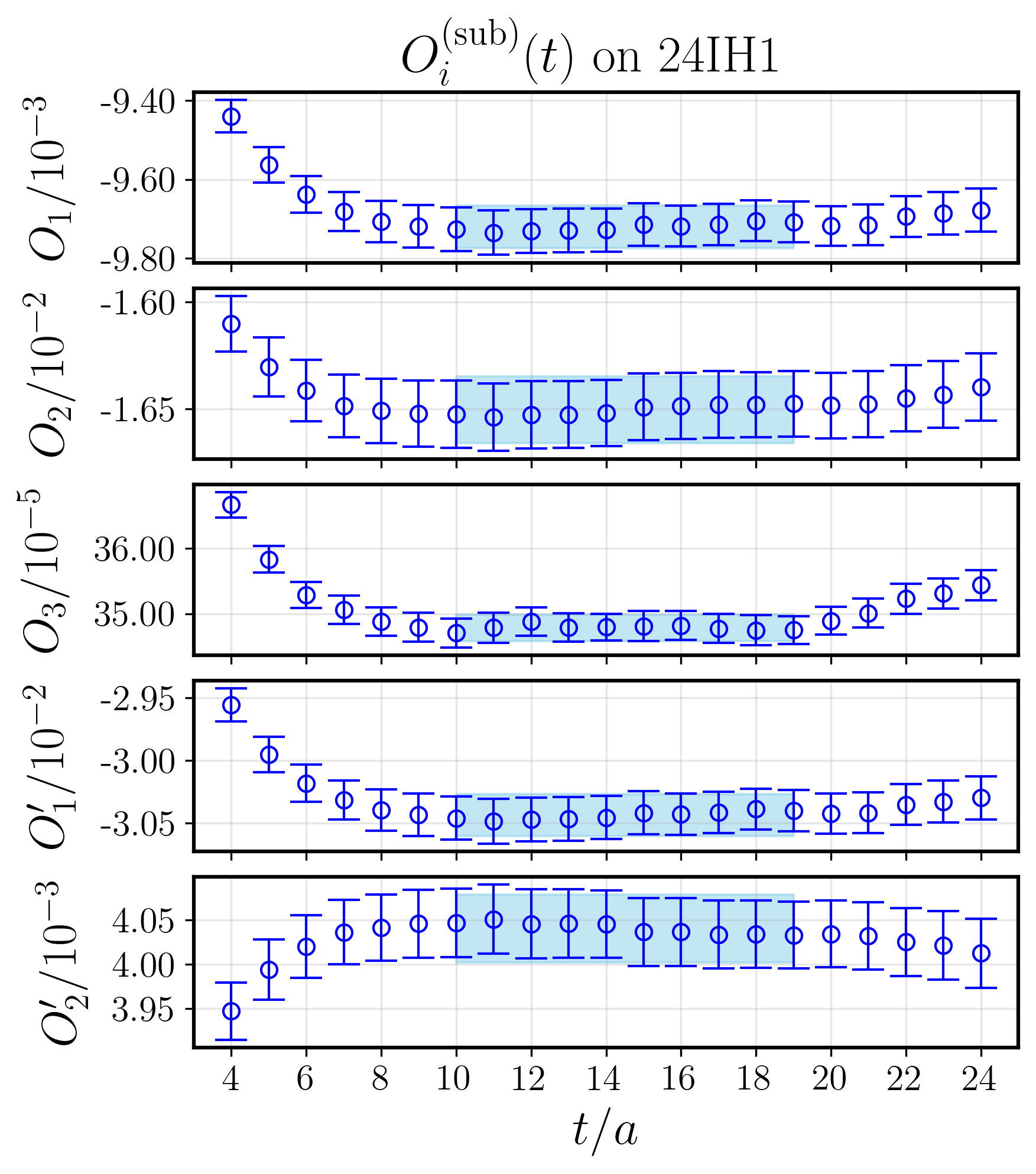}
	\includegraphics[height=0.55\textwidth]{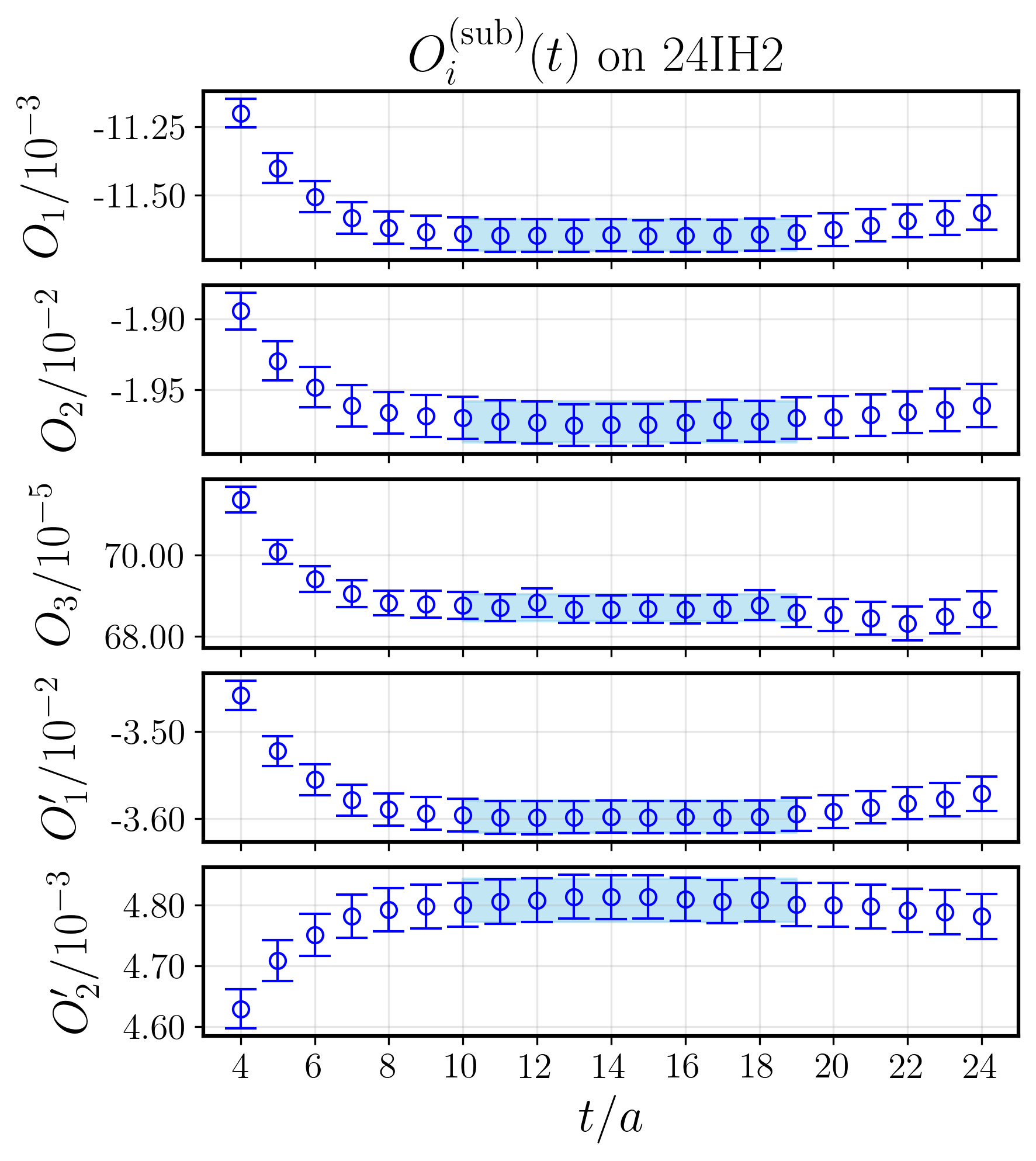}
	\caption{Effective bare matrix elements in the 24IH1 and 24IH2 ensembles. The blue data points represent the values of $O_i^{\text{sub}}(t)$ at different $t$, while the light blue bands indicate the bare matrix elements $\langle \mathcal{O}_i\rangle$ extracted from the plateau regions.
		\label{fig:results1}
	}
\end{figure}

\begin{figure} 
	\centering
	\includegraphics[height=0.55\textwidth]{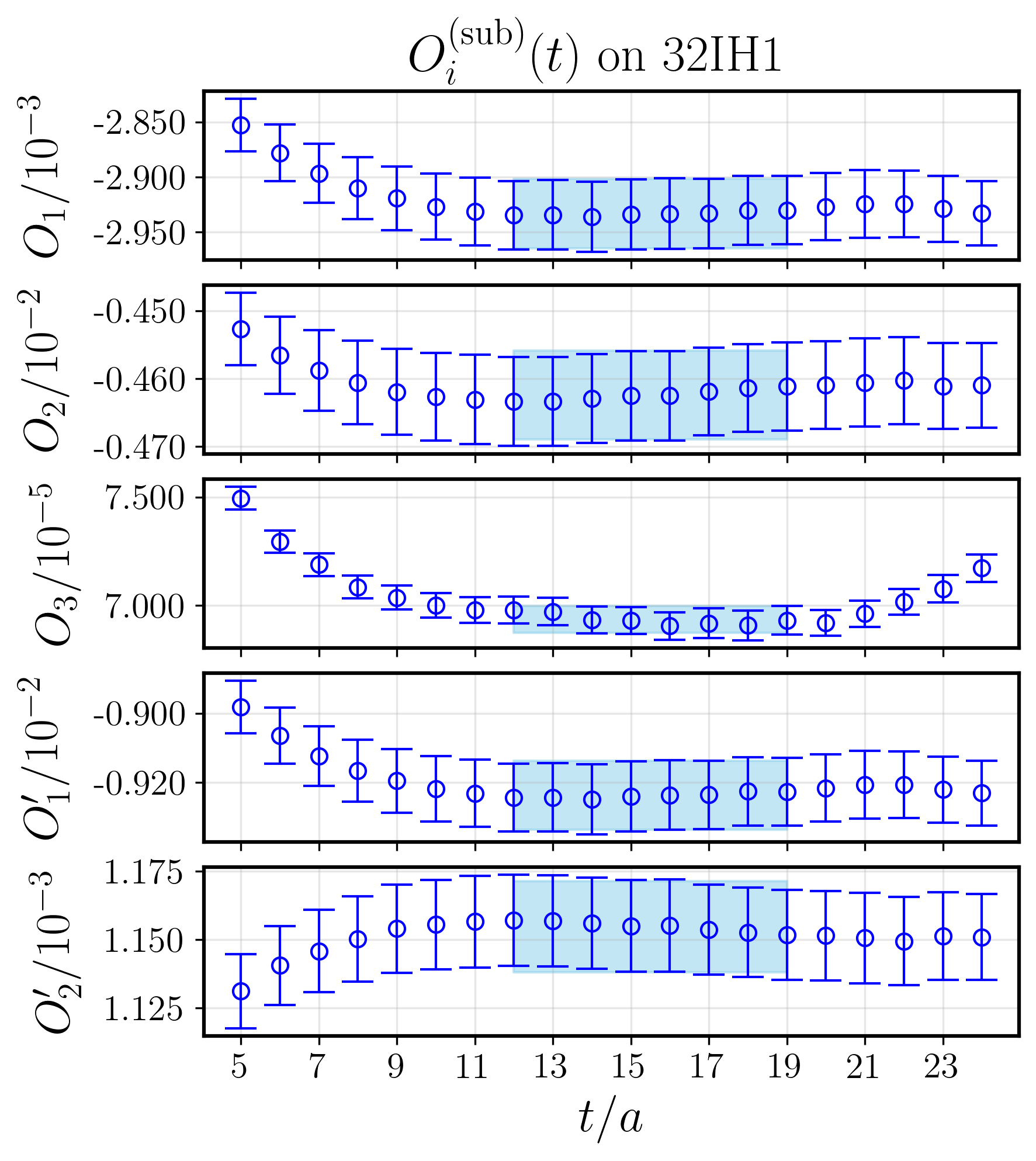}
	\includegraphics[height=0.55\textwidth]{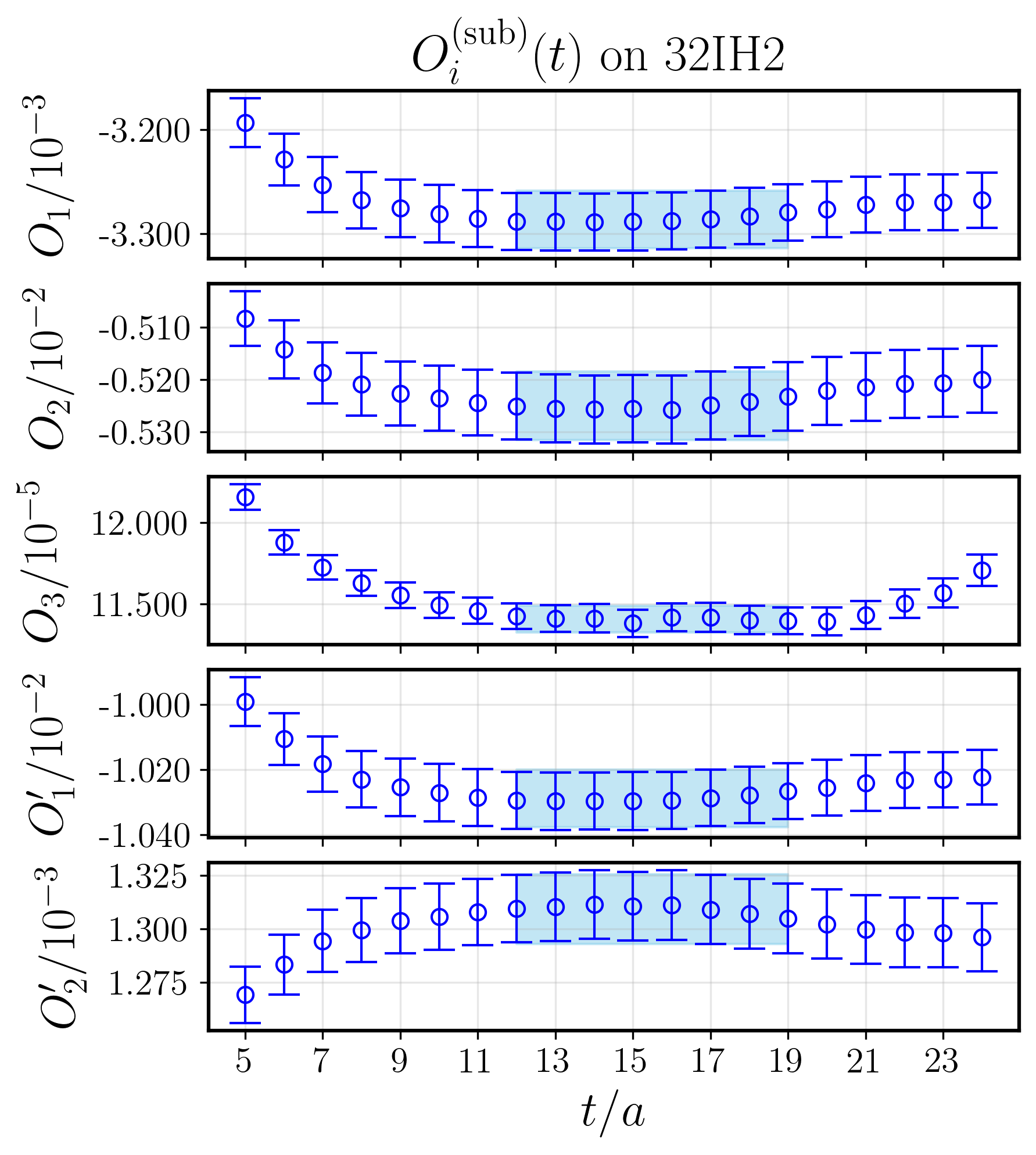}
	\caption{Same as Fig.~\ref{fig:results1}, but with 32IH1 and 32IH2 ensembles.
		\label{fig:results2}
	}
\end{figure}

\begin{figure} 
	\centering
	\includegraphics[height=0.55\textwidth]{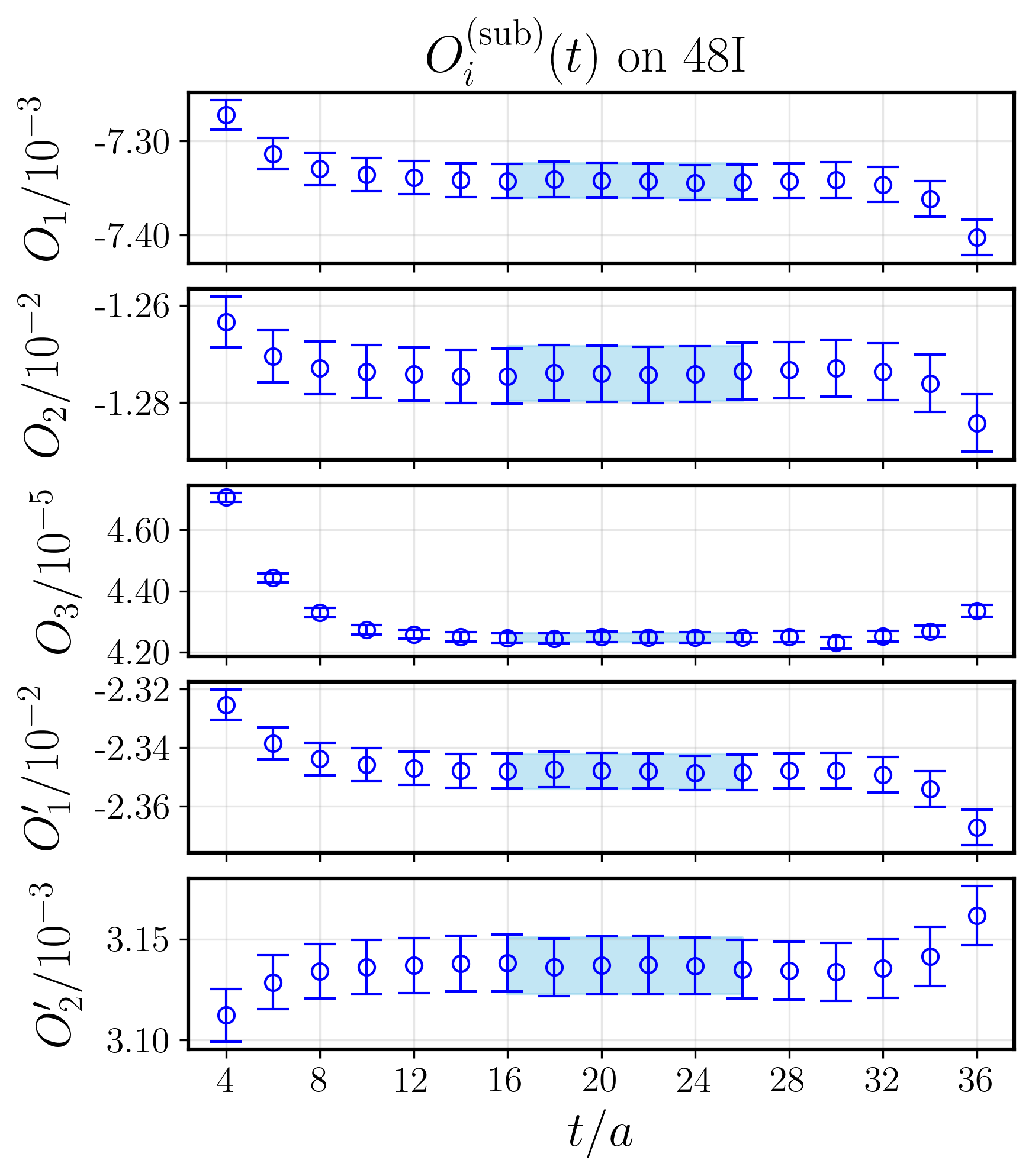}
	\includegraphics[height=0.55\textwidth]{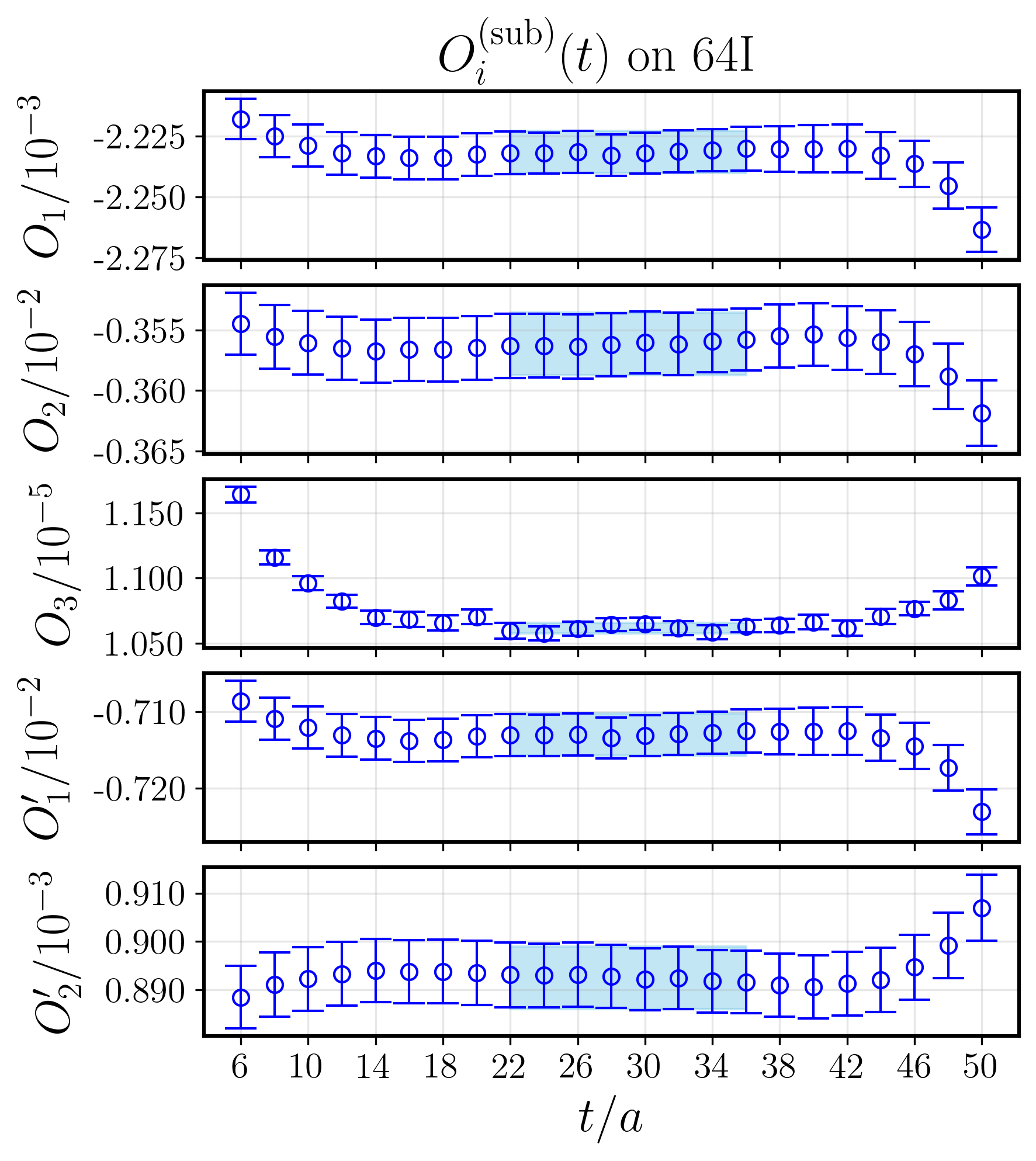}
	\caption{Same as Fig.~\ref{fig:results1}, but with 48I and 64I ensembles.
		\label{fig:results3}
	}
\end{figure}

The results for $O_i^{\text{sub}}(t)$ are shown in Fig.~\ref{fig:results1} (for 24IH1 and 24IH2), Fig.~\ref{fig:results2} (for 32IH1 and 32IH2), and Fig.~\ref{fig:results3} (for 48I and 64I). The blue data points represent the values of $O_i^{\text{sub}}(t)$ at different $t$, while the light blue bands indicate the bare matrix elements $\langle \mathcal{O}_i\rangle$ extracted from the plateau regions. In selecting the plateau region, we ensure that $t$ is sufficiently large to suppress excited-state contributions, yet remains well below the region where residual around-the-world effects (originating from the D diagram in Fig.~\ref{fig:ATW}) become significant. A suitable plateau region is found for all ensembles. In particular, for the physical pion mass ensembles 48I and 64I, we adopt the plateau range $t\in [1.8~\text{fm},3.0~\text{fm}]$. Table~\ref{table:bare_results} summarizes the bare matrix elements obtained from these plateau regions.
\begin{table}
	\centering
	\begin{tabular}{lccccc}
		\hline \hline  &$a^4\langle \mathcal{O}_1\rangle/10^{-3}$&$a^4\langle \mathcal{O}_2\rangle/10^{-3}$&$a^4\langle \mathcal{O}_3\rangle/10^{-5}$&$a^4\langle \mathcal{O}^\prime_{1}\rangle/10^{-3}$&$a^4\langle\mathcal{O}^\prime_{2}\rangle/10^{-3}$ \\
		\hline 
		24IH1 & $-9.721(54)$ & $-16.50(16)$ & $34.78(21)$ & $-30.44(17)$ & $4.041(38)$ \\
		24IH2 & $-11.644(59)$ & $-19.72(15)$ & $68.70(34)$ & $-35.98(19)$ & $4.808(36)$ \\
		32IH1 & $-2.933(32)$ & $-4.624(66)$ & $6.935(62)$ & $-9.24(10)$ & $1.155(17)$ \\
		32IH2 & $-3.286(28)$ & $-5.250(65)$ & $11.408(82)$ & $-10.288(88)$ & $1.309(16)$ \\
		48I & $-7.343(19)$ & $-12.741(58)$ & $4.247(15)$ & $-23.481(60)$ & $3.137(15)$ \\
		64I & $-2.2315(86)$ & $-3.561(26)$ & $1.0615(42)$ & $-7.130(27)$ & $0.8925(65)$ \\
		\hline \hline
	\end{tabular}
	\caption{\label{table:bare_results}Bare matrix elements fitted from the plateau regions shown in Fig.~\ref{fig:results1} (24IH1 and 24IH2), Fig.~\ref{fig:results2} (32IH1 and 32IH2), and Fig.~\ref{fig:results3} (48I and 64I). The results are shown in lattice units.}
\end{table}

The matrix element of $\mathcal{O}_{3}$ is directly related to the bag parameter associated with neutral meson mixing in the Standard Model. Since lattice studies of the bag parameter are well established, they provide a direct cross-check to our results. In Ref.~\cite{Aoki:2010pe}, the same 24I and 32I ensembles were used to compute the bare bag parameter using various valence quark masses for $m_l$ and $m_s$. In particular, when both the light and heavy quark masses are set to $m_{l,\text{sea}}$, the resulting matrix elements can be directly compared with the bare matrix element of $\mathcal{O}_{3}$. Specifically, Tables III and IV in Ref.~\cite{Aoki:2010pe} present the fitted bare matrix element $B^{\text{lat}}(m_x,m_y)$, extracted from the ratio
\begin{equation}
	\frac{\langle P(t_1)| \mathcal{O}_{VV+AA}(t)|P(t_2)\rangle}{\frac{8}{3}\langle P(t_1) | A_0(t)\rangle\langle A_0(t)| P(t_2)\rangle},
\end{equation}
where $P$ denotes a meson consisting of a light quark with mass $m_x$ and a heavy quark with mass $m_y$, and $A_0$ is the time component of the axial-vector current. The four-quark operator associated with the bag parameter is defined as 
$\mathcal{O}_{VV+AA}=(\bar{\psi}_y\gamma_\mu \psi_x)(\bar{\psi}_y\gamma_\mu \psi_x)+(\bar{\psi}_y\gamma_\mu\gamma_5 \psi_x)(\bar{\psi}_y\gamma_\mu\gamma_5 \psi_x)$. When $\psi_x=d$ and $\psi_y=s$ (i.e., $m_x=m_l$ and $m_y=m_s$), the expression above corresponds to the neutral $K$ meson and its bare bag parameter $B_K^{\text{lat}}$. When $\psi_x=d$ and $\psi_y=u$ (i.e., $m_x=m_y=m_l$), the operator $\mathcal{O}_{VV+AA}$ is related to the operator defined in Eq.~(\ref{Operators}) and Eq.~(\ref{QO}) by $\mathcal{O}_{VV+AA}=Q_1=2\mathcal{O}_3$. Consequently, the results in Ref.~\cite{Aoki:2010pe} can be directly compared with our calculation of the bare matrix elements:
\begin{equation}
	B_\pi^{\text{lat}}=B^{\text{lat}}(m_x,m_y)\Big|_{m_x=m_y=m_{l}}=\frac{2a^4\langle \mathcal{O}_{3} \rangle}{\frac{8}{3}(af_\pi/Z_A)^2 (am_\pi)^2},
\end{equation}
where $f_\pi$ is normalized according to the PDG convention with $f_\pi \approx 130~$MeV~\cite{ParticleDataGroup:2024cfk}.

In Table~\ref{table:Bpilat} we present a comparison between our results of \(B_\pi^{\mathrm{lat}}\) and those reported in Ref.~\cite{Aoki:2010pe}. In Ref.~\cite{Aoki:2010pe}, around-the-world effects are removed by combining periodic and antiperiodic temporal boundary conditions. Although our procedure differs, our results agree with those of Ref.~\cite{Aoki:2010pe} within statistical uncertainties. 
We also list the values of \(B_\pi^{\mathrm{lat}}\) converted from the bare matrix elements \(\langle \pi^+|\mathcal{O}_3|\pi^-\rangle\) provided in Ref.~\cite{Detmold:2022jwu}, using their values of \(a m_\pi\), \(a f_\pi\), and \(Z_A\) in Ref.~\cite{Detmold:2020jqv}. The corresponding ratios of our results to theirs are also presented, which are approximately $2$ within statistical errors. We also checked that all five bare matrix elements have a similar discrepancy of a factor of $2$, thus the most likely source of this difference appears to be an overall normalization factor. 
Our normalization convention agrees with that used in Ref.~\cite{Aoki:2010pe}, which, in the case of $\psi_y=s$, yields a correct \(B_K^{\overline{\mathrm{MS}}}(\mu)\) result that has been included in the FLAG review~\cite{FlavourLatticeAveragingGroupFLAG:2024oxs}. Thus, this provides support for the correctness of the normalization used in our work.

\begin{table}
	\centering
	\begin{tabular}{ccccc}
		\hline \hline 
		$B_\pi^{\text{lat}}$ & 24IH1 & 24IH2 & 32IH1 & 32IH2 \\
		\hline 
		This work & $0.5072(30)$ & $0.5433(26)$ & $0.4691(42)$ & $0.4963(36)$ \\
		Ref.~\cite{Aoki:2010pe} & $0.5075(31)$ & $0.5439(18)$ & $0.4721(26)$ & $0.5004(22)$ \\
		Ref.~\cite{Detmold:2022jwu} & $0.2544(21)$ & $0.2705(13)$ & $0.2299(27)$ & $0.2445(18)$ \\
        \hline
        This work/Ref.~\cite{Detmold:2022jwu} & $1.994(20)$ & $2.008(14)$ & $2.041(30)$ & $2.030(21)$ \\
		\hline \hline
	\end{tabular}
	\caption{\label{table:Bpilat}     
    Computed values of the bare matrix element $B_\pi^{\text{lat}}$ in the 24IH1, 24IH2, 32IH1, and 32IH2 ensembles. The first row shows our results, the second row shows the values reported in Tables III and IV of Ref.~\cite{Aoki:2010pe}. The third row presents the results computed from the matrix elements \(\langle \pi^+|\mathcal{O}_3|\pi^-\rangle\) provided in Ref.~\cite{Detmold:2022jwu}, using their values of \(a m_\pi\), \(a f_\pi\), and \(Z_A\) in Ref.~\cite{Detmold:2020jqv}. The fourth row shows the ratios of our results to those computed from Ref.~\cite{Detmold:2022jwu}.}
\end{table}

\subsection{Renormalization and Continuum Extrapolation}
\begin{figure} 
	\centering
	\includegraphics[height=0.7\textwidth]{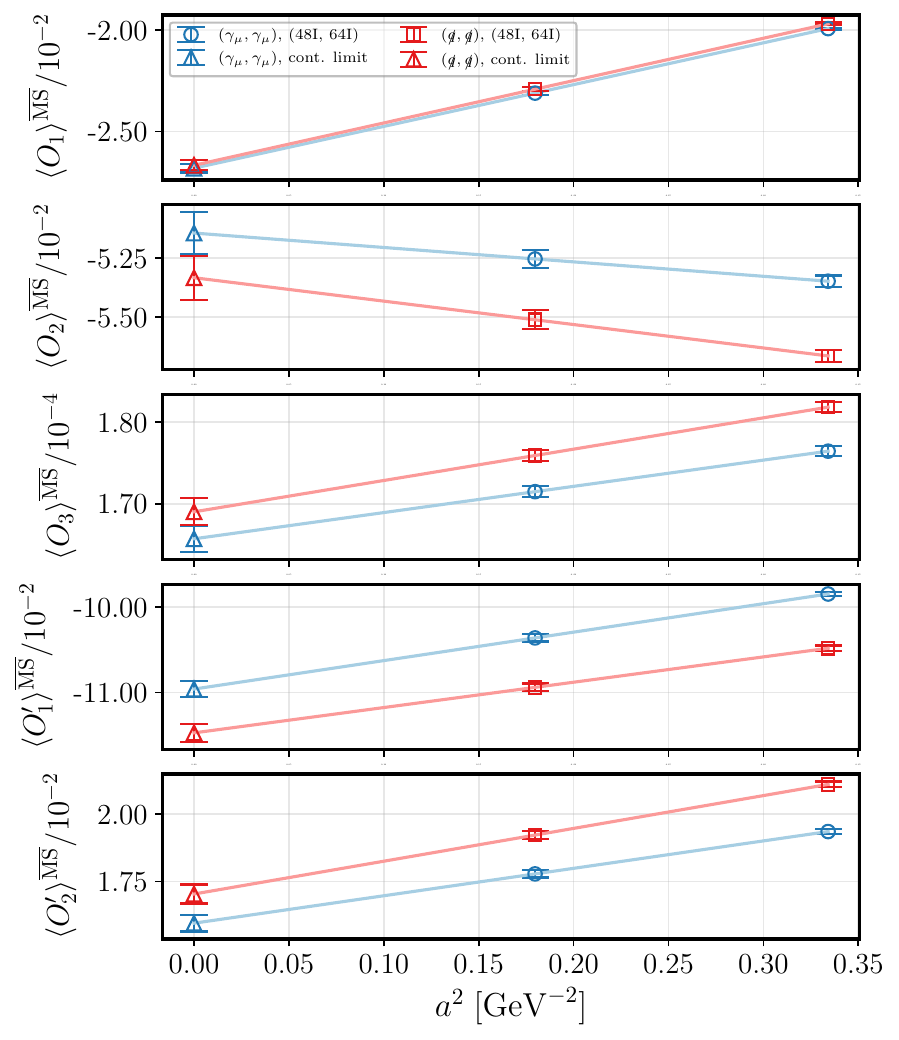}
	\caption{Renormalized matrix elements $\langle\mathcal{O}_i\rangle(\mu=3\,\text{GeV})$ [GeV$^4$] computed using both the $(\gamma_\mu,\gamma_\mu)$ and $(\slashed{q},\slashed{q})$ schemes for the 48I and 64I ensembles, together with the corresponding continuum-extrapolated results. The difference between results in $(\gamma_\mu,\gamma_\mu)$ and $(\slashed{q},\slashed{q})$ schemes can be viewed as an estimate of perturbative truncation errors.
		\label{fig:cont_lim}
	}
\end{figure}

The renormalized matrix elements are obtained by multiplying the bare results by the renormalization factors given in Eq.~\eqref{NPR} and Appendix~\ref{Appendix:ZRI}. Tables~\ref{table: ren1} and~\ref{table: ren2} summarize the renormalized matrix elements computed using the $(\gamma_\mu,\gamma_\mu)$ and $(\slashed{q},\slashed{q})$ schemes, together with the continuum extrapolation results from the physical pion mass ensembles. To avoid uncertainties associated with the chiral extrapolation, we perform a direct continuum extrapolation using the physical pion mass ensembles 48I and 64I, as shown in Fig.~\ref{fig:cont_lim}. 
\begin{table}
	\centering
	\begin{tabular}{c|c|ccccc}
		\hline \hline 
        & Ensemble & $\left\langle\mathcal{O}_1\right\rangle^{\overline{\mathrm{MS}}} / 10^{-2}$ & $\left\langle\mathcal{O}_2\right\rangle^{\overline{\mathrm{MS}}} / 10^{-2}$ & $\left\langle\mathcal{O}_3\right\rangle^{\overline{\mathrm{MS}}} / 10^{-4}$ & $\left\langle\mathcal{O}^\prime_{1}\right\rangle^{\overline{\mathrm{MS}}} / 10^{-2}$ & $\left\langle\mathcal{O}^\prime_{2}\right\rangle^{\overline{\mathrm{MS}}} / 10^{-2}$ \\
		\hline 
        \multirow{4}{*}{Unphysical $m_\pi$} 
        & 24IH1 & $-3.051(18)$ & $-7.823(74)$ & $16.51(10)$ & $-14.39(8)$ & $2.796(27)$ \\
		& 24IH2 & $-3.690(19)$ & $-9.349(69)$ & $32.60(16)$ & $-17.00(9)$ & $3.329(25)$ \\
		& 32IH1 & $-3.189(35)$ & $-7.095(101)$ & $11.69(11)$ & $-13.95(16)$ & $2.396(35)$ \\
		& 32IH2 & $-3.588(30)$ & $-8.055(101)$ & $19.23(14)$ & $-15.54(14)$ & $2.716(35)$ \\
		\hline 
        \multirow{3}{*}{Physical $m_\pi$} 
        & 48I & $-1.993(6)$ & $-5.347(25)$ & $1.765(7)$ & $-9.849(27)$ & $1.935(9)$ \\
		& 64I & $-2.310(10)$ & $-5.253(39)$ & $1.715(7)$ & $-10.362(42)$ & $1.778(14)$ \\
		& Cont. lim. & $-2.680(22)$ & $-5.143(89)$ & $1.657(17)$ & $-10.960(95)$ & $1.596(31)$ \\
		\hline \hline
	\end{tabular}
	\caption{Renormalized matrix elements $\langle\mathcal{O}_i\rangle^{\overline{\mathrm{MS}}}(\mu=3~\text{GeV})$ [GeV$^4$] in the $(\gamma_\mu,\gamma_\mu)$ scheme. The upper panel shows results for ensembles with unphysical pion masses (24IH1, 24IH2, 32IH1, and 32IH2), whereas the lower panel presents results at the physical pion mass (48I and 64I). Additionally, the continuum-extrapolated results with only statistical errors are shown. \label{table: ren1}}
\end{table}

\begin{table}
	\centering
	\begin{tabular}{c|c|ccccc}
		\hline \hline 
        & Ensemble & $\left\langle\mathcal{O}_1\right\rangle^{\overline{\mathrm{MS}}} / 10^{-2}$ & $\left\langle\mathcal{O}_2\right\rangle^{\overline{\mathrm{MS}}} / 10^{-2}$ & $\left\langle\mathcal{O}_3\right\rangle^{\overline{\mathrm{MS}}} / 10^{-4}$ & $\left\langle\mathcal{O}^\prime_{1}\right\rangle^{\overline{\mathrm{MS}}} / 10^{-2}$ & $\left\langle\mathcal{O}^\prime_{2}\right\rangle^{\overline{\mathrm{MS}}} / 10^{-2}$ \\
		\hline 
        \multirow{4}{*}{Unphysical $m_\pi$} 
        & 24IH1 & $-3.010(17)$ & $-8.294(83)$ & $17.01(10)$ & $-15.33(9)$ & $3.054(33)$ \\
		& 24IH2 & $-3.642(19)$ & $-9.912(80)$ & $33.59(17)$ & $-18.10(10)$ & $3.636(32)$ \\
		& 32IH1 & $-3.160(34)$ & $-7.467(106)$ & $12.01(11)$ & $-14.78(16)$ & $2.598(39)$ \\
		& 32IH2 & $-3.556(30)$ & $-8.478(107)$ & $19.76(15)$ & $-16.46(15)$ & $2.945(38)$ \\
		\hline 
        \multirow{3}{*}{Physical $m_\pi$} 
        & 48I & $-1.970(7)$ & $-5.665(27)$ & $1.819(7)$ & $-10.483(32)$ & $2.110(11)$ \\
		& 64I & $-2.291(11)$ & $-5.511(41)$ & $1.759(7)$ & $-10.941(45)$ & $1.922(16)$ \\
		& Cont. lim. & $-2.666(25)$ & $-5.333(95)$ & $1.690(17)$ & $-11.473(104)$ & $1.704(36)$ \\
		\hline \hline
	\end{tabular}
	\caption{Same as Table.~\ref{table: ren1}, but with $(\slashed{q},\slashed{q})$ scheme.\label{table: ren2}}
\end{table}

Our final results are obtained in the $(\gamma_\mu,\gamma_\mu)$ scheme and we treat the difference between the $(\gamma_\mu,\gamma_\mu)$ and $(\slashed{q},\slashed{q})$ schemes as a systematic uncertainty arising from the perturbative truncation of the matching from $\mathrm{RI}$ to $\overline{\mathrm{MS}}$ ($R_{ij}^{\overline{\mathrm{MS}}\leftarrow \mathrm{RI}}(\mu)$ in Eq.~(\ref{NPR})). To estimate finite-volume effects, we follow the approach described in Ref.~\cite{Nicholson:2018mwc}, where tadpole integrals in $\chi$EFT are replaced by discrete momentum sums in a finite volume. These effects are smaller than the statistical uncertainties, so we incorporate them only in our error budget. The final results for the renormalized matrix elements $\left\langle\mathcal{O}_i\right\rangle^{\overline{\mathrm{MS}}}$ and the bag parameter \(B_\pi^{\overline{\rm MS}}(\mu) =2\langle \mathcal{O}_3\rangle^{\overline{\rm MS}}(\mu)/(\frac{8}{3} m_\pi^2 f_\pi^2)\) are presented in Table~\ref{table:results}. The statistical correlation matrix for our final results $\langle\mathcal{O}_i\rangle^{\overline{\mathrm{MS}}}$ is presented in Table~\ref{table:corr_mat}.

For reference, we include the results from Refs.~\cite{Nicholson:2018mwc,Detmold:2022jwu} in Table~\ref{table:results}; note that we also present $B_\pi^{\overline{\rm MS}}(\mu)$ converted from the matrix element $\langle \pi^+|\mathcal{O}_3(\mu)|\pi^-\rangle^{\overline{\mathrm{MS}}}$ provided in Ref.~\cite{Detmold:2022jwu}. Comparing with these previous works, the uncertainties are reduced in our results. This improvement is due to the improved statistical precision and the avoidance of systematic uncertainties introduced by the chiral extrapolation. Furthermore, we perform a more reliable estimation of perturbative truncation effects by applying both $(\gamma_\mu,\gamma_\mu)$ and
$(\slashed{q},\slashed{q})$ schemes. The ratios of our results to those of Ref.~\cite{Detmold:2022jwu} are consistent with $2$ within errors, suggesting a difference of an overall normalization factor. In comparison with Ref.~\cite{Nicholson:2018mwc}, our results exhibit discrepancies—approximately $1.8\sigma$ for $\langle\mathcal{O}_3\rangle^{\overline{\mathrm{MS}}}$, and $3\sigma\sim 6\sigma$ for the other matrix elements. Since $\langle\mathcal{O}_3\rangle^{\overline{\mathrm{MS}}}$ does not mix with other operators under renormalization, we speculate that these discrepancies may be related to the nonperturbative renormalization procedure. Additionally, lattice artifacts may also play a role in these differences, as these matrix elements have mass dimension four and are therefore sensitive to discretization effects.
\begin{table}
	\centering
	\begin{tabular}{ccccc}
		\hline \hline Operators & This work & Ref.~\cite{Nicholson:2018mwc} & Ref.~\cite{Detmold:2022jwu} &This work/Ref.~\cite{Detmold:2022jwu}\\
		\hline
		 $\left\langle\mathcal{O}_1\right\rangle^{\overline{\mathrm{MS}}} / 10^{-2} $ [$\text{GeV}^4$] & $-2.680(22)_{\text{stat}}(14)_{\text{PT}}(20)_{L}$ & $-1.89(13)$ & $-1.27(16)$ &$2.11(27)$\\
		 $\left\langle\mathcal{O}_2\right\rangle^{\overline{\mathrm{MS}}} / 10^{-2}$ [$\text{GeV}^4$] & $-5.143(89)_{\text{stat}}(190)_{\text{PT}}(39)_{L}$ & $-3.77(32)$ & $-2.45(22)$&$2.10(21)$\\
		 $\left\langle\mathcal{O}_3\right\rangle^{\overline{\mathrm{MS}}} / 10^{-4}$ [$\text{GeV}^4$] & $1.657(17)_{\text{stat}}(33)_{\text{PT}}(0)_{L}$ & $1.86(10)$ & $0.869(80)$&$1.91(18)$\\
		 $\left\langle\mathcal{O}^\prime_1\right\rangle^{\overline{\mathrm{MS}}} / 10^{-2}$ [$\text{GeV}^4$] & $-10.96(10)_{\text{stat}}(52)_{\text{PT}}(8)_{L}$ & $-7.81(54)$ & $-5.35(48)$&$2.05(21)$\\
		 $\left\langle\mathcal{O}^\prime_2\right\rangle^{\overline{\mathrm{MS}}} / 10^{-2}$ [$\text{GeV}^4$] & $1.596(31)_{\text{stat}}(108)_{\text{PT}}(12)_{L}$ & $1.23(11)$ & $0.757(75)$&$2.11(26)$\\
		\hline
		$B_\pi^{\overline{\rm{MS}}}$ & $0.3769(24)_{\text{stat}}(76)_{\text{PT}}(1)_{L}$ & $0.421(23)$ & $0.197(18)$ &$1.91(18)$ \\
		\hline \hline
	\end{tabular}
	\caption{Final results for the renormalized matrix elements $\langle\mathcal{O}_i\rangle^{\overline{\mathrm{MS}}}(\mu=3\,\text{GeV})$ and the bag parameter $B_\pi^{\overline{\rm MS}}(\mu=3\,\text{GeV})$. We include statistical, perturbative truncation and finite-volume uncertainties in our error budget, denoted by the subscripts “stat”, “PT”, and “L”, respectively. The perturbative truncation errors associated with the matching from $\mathrm{RI}$ to $\overline{\mathrm{MS}}$ scheme are estimated by taking the difference between the $(\gamma_\mu,\gamma_\mu)$ and $(\slashed{q},\slashed{q})$ schemes. Finite-volume effects are estimated using the formula provided in Ref.~\cite{Nicholson:2018mwc}. For comparison, we list results from Refs.~\cite{Nicholson:2018mwc,Detmold:2022jwu}. We also show the ratios of our results to those from Refs.~\cite{Detmold:2022jwu}.\label{table:results}}
\end{table}

\begin{table}
	\centering
\begin{tabular}{||c||c|c|c|c|c||}
\hline & $\left\langle\mathcal{O}_1\right\rangle^{\overline{\mathrm{MS}}}$ & $\left\langle\mathcal{O}_2\right\rangle^{\overline{\mathrm{MS}}}$ & $\left\langle\mathcal{O}_3\right\rangle^{\overline{\mathrm{MS}}}$ & $\left\langle\mathcal{O}^\prime_1\right\rangle^{\overline{\mathrm{MS}}}$ &$\left\langle\mathcal{O}^\prime_2\right\rangle^{\overline{\mathrm{MS}}}$ \\
\hline \hline$\left\langle\mathcal{O}_1\right\rangle^{\overline{\mathrm{MS}}}$ & 1.0 & 0.609 & -0.517 & 0.907 & -0.595 \\
\hline$\left\langle\mathcal{O}_2\right\rangle^{\overline{\mathrm{MS}}}$ & 0.609 & 1.0 & -0.610 & 0.590& -0.987\\
\hline$\left\langle\mathcal{O}_3\right\rangle^{\overline{\mathrm{MS}}}$ & -0.517 & -0.610 & 1.0 & -0.498& 0.602\\
\hline$\left\langle\mathcal{O}^\prime_1\right\rangle^{\overline{\mathrm{MS}}}$ & 0.907 & 0.590 & -0.498 & 1.0 & -0.575\\
\hline$\left\langle\mathcal{O}^\prime_2\right\rangle^{\overline{\mathrm{MS}}}$ & -0.595 & -0.987 & 0.602 & -0.575 & 1.0\\
\hline \hline
\end{tabular}
\caption{The statistical correlation matrix for the renormalized matrix elements $\langle\mathcal{O}_i\rangle^{\overline{\mathrm{MS}}}(\mu=3\,\text{GeV})$ in the $(\gamma_\mu,\gamma_\mu)$ scheme, estimated using bootstrap samples. It reveals strong statistical correlations within each chirally mixed operator pair: ($\mathcal{O}_1$, $\mathcal{O}^\prime_1$) and ($\mathcal{O}_2$, $\mathcal{O}^\prime_2$).
\label{table:corr_mat}}
\end{table}

\section{Conclusion\label{sec4}}
Short-range matrix elements associated with the $\pi^-\to\pi^+ ee$ process arise at leading order in the $0\nu\beta\beta$ decay channel $nn \to pp ee$. 
However, previous lattice calculations \cite{Nicholson:2018mwc,Detmold:2022jwu} exhibit significant discrepancies in their numerical results, prompting the need for an independent lattice calculation as a cross-check. 
We employed domain wall fermion ensembles generated by the RBC/UKQCD Collaboration at the physical pion mass to calculate both the bare and renormalized matrix elements. 
To address systematic uncertainties introduced by around-the-world effects, we proposed a subtraction method that reconstructs and removes these effects directly from the lattice data, thereby achieving stable plateaus in the ratio of three-point to two-point correlation functions. 
Next, we perform a nonperturbative renormalization of the four-quark operators using the RI/SMOM method in \((\gamma_\mu, \gamma_\mu)\) and \((\slashed{q}, \slashed{q})\) schemes. 
We estimate the systematic error from perturbative truncation of the matching from $\mathrm{RI}$ to $\overline{\mathrm{MS}}$ scheme by examining the difference between these two renormalization schemes. 
Finally, we perform a continuum extrapolation using two physical-pion-mass ensembles with similar volume but different lattice spacings, obtaining the short-range matrix elements.

Compared with previous work, our calculation is the first to determine these matrix elements directly at physical quark masses with a continuum extrapolation.
The uncertainties in our matrix element calculations are significantly reduced. For all five matrix elements, we find that the ratios of our bare matrix elements to those in Ref.~\cite{Detmold:2022jwu} are statistically consistent with 2, suggesting a difference of an overall normalization factor but otherwise good agreement on individual data points if this postulated factor of 2 is taken into account. The correctness of the normalization used in our work is supported by agreement with Ref.~\cite{Aoki:2010pe}, which yields a reliable determination of \(B^{\overline{\mathrm{MS}}}_K(\mu)\) that is included in the FLAG review~\cite{FlavourLatticeAveragingGroupFLAG:2024oxs} and consistent with the world average. Our results still differ from those of Ref.~\cite{Nicholson:2018mwc} by $2\sigma$–$6\sigma$, depending on the matrix element.

Future improvements in precision may be achieved through finer lattice spacings, increased statistics, and higher-order perturbative matching of nonperturbative lattice matrix elements to the $\overline{\mathrm{MS}}$ scheme, such as the one recently performed for the case of the kaon bag parameters~\cite{Gorbahn:2024qpe}. 
Additionally, in $\chi$EFT, four-nucleon contact interactions can also contribute at leading order through the renormalization procedure~\cite{Cirigliano:2018hja}. 
Hence, extending lattice QCD methods to compute analogous short-range matrix elements involving nucleons is an important future research direction. 
Such progress would reduce nonperturbative QCD uncertainties in theoretical calculations and, through $0\nu\beta\beta$ experiments, shed light on the nature of neutrino masses, the violation of lepton number, and the matter–antimatter asymmetry of the universe.

\acknowledgements
We would like to thank our RBC and UKQCD Collaboration colleagues for helpful discussions and support. 
P.B. and T.I. were supported in part by US DOE Contract DESC0012704(BNL) and the Scientific Discovery
through Advanced Computing (SciDAC) program LAB
22-2580. F.E. has received funding from the European Union's Horizon Europe research and innovation programme under the Marie Sk\l{}odowska-Curie grant agreement No.~101106913. X.F. has been supported in part by NSFC of China under Grant No. 12125501. 
L.J. acknowledge the support of DOE Office of Science Early Career Award DE-SC0021147 and DOE
grant DE-SC0010339. X.Y.T has been supported by US DOE Contract
DESC0012704(BNL). N.G. acknowledges support from STFC grant ST/X000699/1.

\appendix
\section{Renormalization Coefficients\label{Appendix:ZRI}}
In Tables~\ref{table:ZRI} and~\ref{table:ZRI2}, we present the renormalization coefficients \(Z_{ij}^{\mathrm{RI}}(\mu,a)/Z_A^2\) in the NPR basis, evaluated at \(\mu = 3.0\,\mathrm{GeV}\), for both the \(\bigl(\gamma_\mu,\gamma_\mu\bigr)\) and \(\bigl(\slashed{q},\slashed{q}\bigr)\) schemes. Chirally forbidden elements are omitted. The four columns correspond to different lattice spacings for the ensembles 48I, 24IH, 64I, and 32IH. 

These renormalization coefficients were originally computed in Ref.~\cite{Boyle:2024gge}. Unlike Ref.~\cite{Boyle:2024gge}, which adopts the SUSY basis, here the renormalization coefficients are presented in the NPR basis. As a result, \(Z_{11}^{\mathrm{RI}}(\mu,a)/Z_A^2\) is identical to that in Ref.~\cite{Boyle:2024gge}, whereas the remaining coefficients are related to those in Ref.~\cite{Boyle:2024gge} by a basis rotation.

\begin{table} 
	\centering
	\begin{tabular}{c|cccc}
		\hline \hline
		\(a^{-1}[\mathrm{GeV}]\) & 1.7295(38) & 1.7848(50) & 2.3586(70) & 2.3833(86)\\
		\hline
		\(Z_{11}/Z_A^2\) & 0.91427(17) & 0.91641(55) & 0.94123(17) & 0.94044(67) \\
		\hline
		\(Z_{22}/Z_A^2\) & 1.049689(94) & 1.04914(48) & 1.04603(12) & 1.04624(22)\\
		\(Z_{23}/Z_A^2\) & 0.27730(16) & 0.27739(41) & 0.27193(45) & 0.27407(18)\\
		\(Z_{32}/Z_A^2\) & 0.038071(83) & 0.036670(72) & 0.025393(59) & 0.025256(37)\\
		\(Z_{33}/Z_A^2\) & 0.88137(73) & 0.87176(34) & 0.80313(98) & 0.79947(85)\\
		\hline
		\(Z_{44}/Z_A^2\) & 0.92584(70) & 0.916862(98) & 0.85458(82) & 0.85113(68)\\
		\(Z_{45}/Z_A^2\) & -0.037741(94) & -0.036173(87) & -0.022771(84) & -0.022597(76)\\
		\(Z_{54}/Z_A^2\) & -0.25049(21) & -0.25088(33) & -0.24830(46) & -0.25083(23) \\
		\(Z_{55}/Z_A^2\) & 1.03765(51) & 1.0424(15) & 1.08570(83) & 1.0873(18)\\
		\hline \hline
	\end{tabular}
	\caption{\label{table:ZRI}Renormalization coefficients 
	\(\displaystyle Z_{ij}^{\mathrm{RI}}(\mu=3.0\,\mathrm{GeV},a)/Z_A^2\) 
	in the \(\bigl(\gamma_\mu,\gamma_\mu\bigr)\) scheme.}
\end{table}

\begin{table} 
	\centering
	\begin{tabular}{c|cccc}
		\hline \hline
		\(a^{-1}[\mathrm{GeV}]\) & 1.7295(38) & 1.7848(50) & 2.3586(70) & 2.3833(86)\\
		\hline
		\(Z_{11}/Z_A^2\) & 0.95466(16) & 0.956518(75) & 0.97828(27) & 0.97895(22)\\
		\hline
		\(Z_{22}/Z_A^2\) & 1.04998(23) & 1.05017(34) & 1.04746(20) & 1.04801(30)\\
		\(Z_{23}/Z_A^2\) & 0.28138(75) & 0.28278(34) & 0.27553(95) & 0.27822(14)\\
		\(Z_{32}/Z_A^2\) & 0.06283(35) & 0.061656(51) & 0.04733(32) & 0.047862(33)\\
		\(Z_{33}/Z_A^2\) & 1.0001(16) & 0.99033(46) & 0.9041(13) & 0.90350(88)\\
		\hline
		\(Z_{44}/Z_A^2\) & 1.0377(16) & 1.0286(33) & 0.9485(13) & 0.94765(72)\\
		\(Z_{45}/Z_A^2\) & -0.04224(29) & -0.040792(46) & -0.02524(18) & -0.02522(10)\\
		\(Z_{54}/Z_A^2\) & -0.28078(71) & -0.280612(90) & -0.2760(11) & -0.27941(22) \\
		\(Z_{55}/Z_A^2\) & 1.1632(11) & 1.1668(23) & 1.2060(16) & 1.2108(33)\\
		\hline \hline
	\end{tabular}
	\caption{\label{table:ZRI2}Renormalization coefficients 
	\(\displaystyle Z_{ij}^{\mathrm{RI}}(\mu=3.0\,\mathrm{GeV},a)/Z_A^2\) 
	in the \(\bigl(\slashed{q},\slashed{q}\bigr)\) scheme.}
\end{table}

\section{Contributions of Individual Diagrams in Around-the-world Effects\label{Append:ATW}}
To further investigate the contributions of diagrams A, B, C, and D in Fig.~\ref{fig:ATW} and to explain why the $R_2$ method effectively suppresses around-the-world effects for all matrix elements except $\mathcal{O}_3$, we determine the coefficients $\mathcal{N}_{A,i}$, $\mathcal{N}_{B,i}$, $\mathcal{N}_{C,i}$, and $\mathcal{N}_{D,i}$ in Eq.~\eqref{OR1} by directly fitting the lattice data. 

To extract the around-the-world effects (i.e., contributions from diagrams B, C, and D in Fig.~\ref{fig:ATW}) from the data, we perform a direct fit to the three-point function $C^i_3(t_1,t_2)$. We define $\Delta T = \max\{t_\pi, t_{\pi\pi}\}$ as the minimal time separation required for the pion and $\pi\pi$ ground states to dominate. If the separations between $\phi_\pi(t_1)$, $\phi_\pi(t_2)$, and $\mathcal{O}_i$ are all greater than $\Delta T$, the three-point function can be well described as a sum of contributions from diagrams A, B, C, and D.

For clarity, we define the temporal separation between two operators $A$ and $B$ as $d(A,B)$, which is taken as the shortest time interval (under periodic boundary conditions) satisfying $d(A,B) \leq T/2$. We introduce the minimal separation among all three operators as:
\begin{equation}
	d_{\min} \equiv \min\{d(\phi_{\pi,1},\phi_{\pi,2}),d(\phi_{\pi,1},\mathcal{O}_i),d(\phi_{\pi,2},\mathcal{O}_i)\}.
\end{equation}
Under the condition $d_{\min} \geq \Delta T$, the three-point function $C^i_3(t_1,t_2)$ can be approximated as
\begin{equation}
\begin{aligned}
	C^i_3(t_1,t_2)\Big|_{d_{\min}\geq \Delta T} &=N_\pi^2\mathcal{N}_{A,i}e^{-m_\pi(t_1+t_2)}+N_\pi^2\mathcal{N}_{B/C,i}e^{-m_\pi(T-t_1-t_2)}(e^{-E_{\pi\pi}t_1}+e^{-E_{\pi\pi}t_2})\\
		&\quad+N_\pi^2\mathcal{N}_{D,i}e^{-m_\pi(t_1+t_2)-E_{\pi\pi}(T-t_1-t_2)}.
\end{aligned}
\end{equation}

\begin{figure} 
	\centering
	\includegraphics[height=0.59\textwidth]{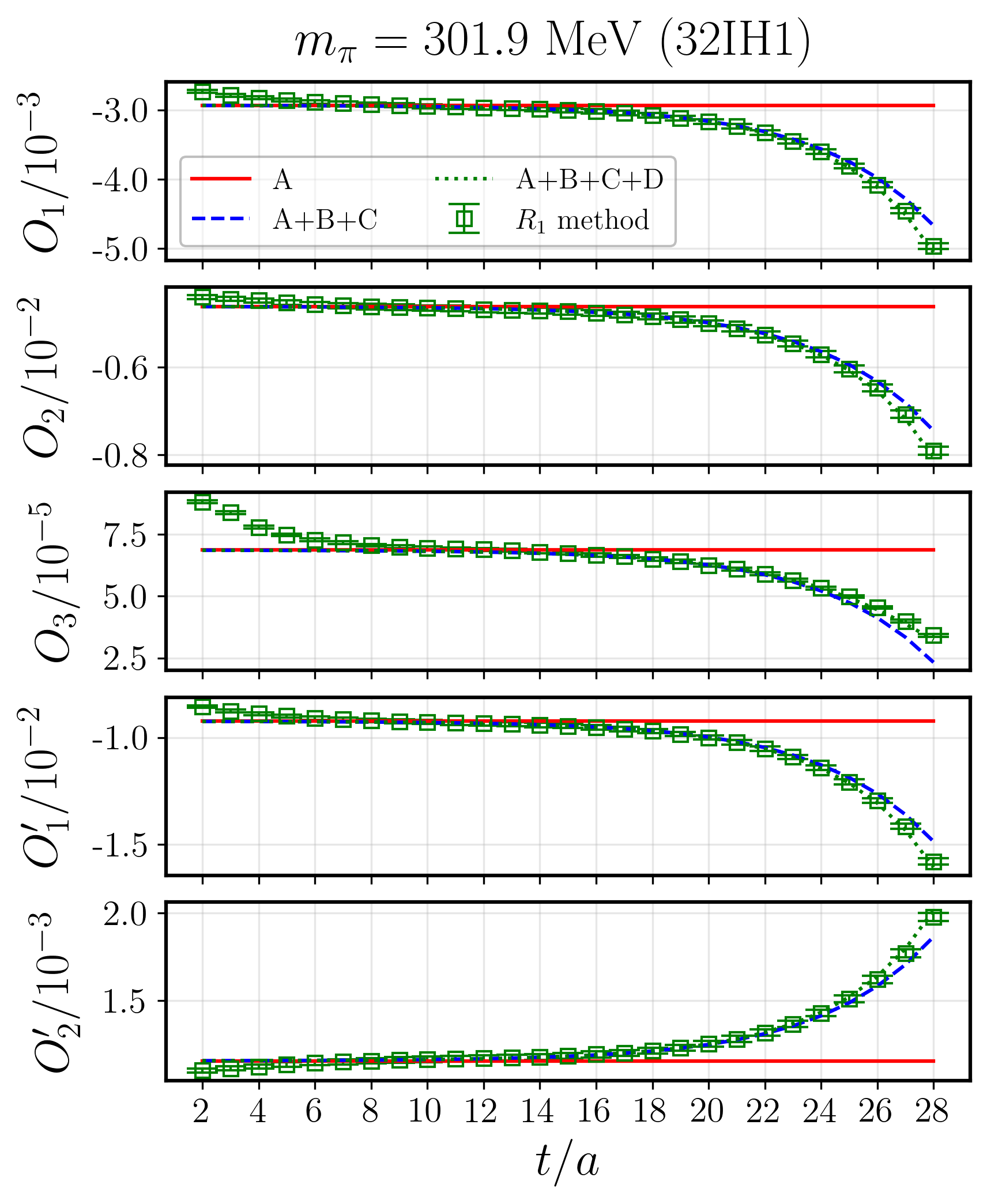}
	\includegraphics[height=0.59\textwidth]{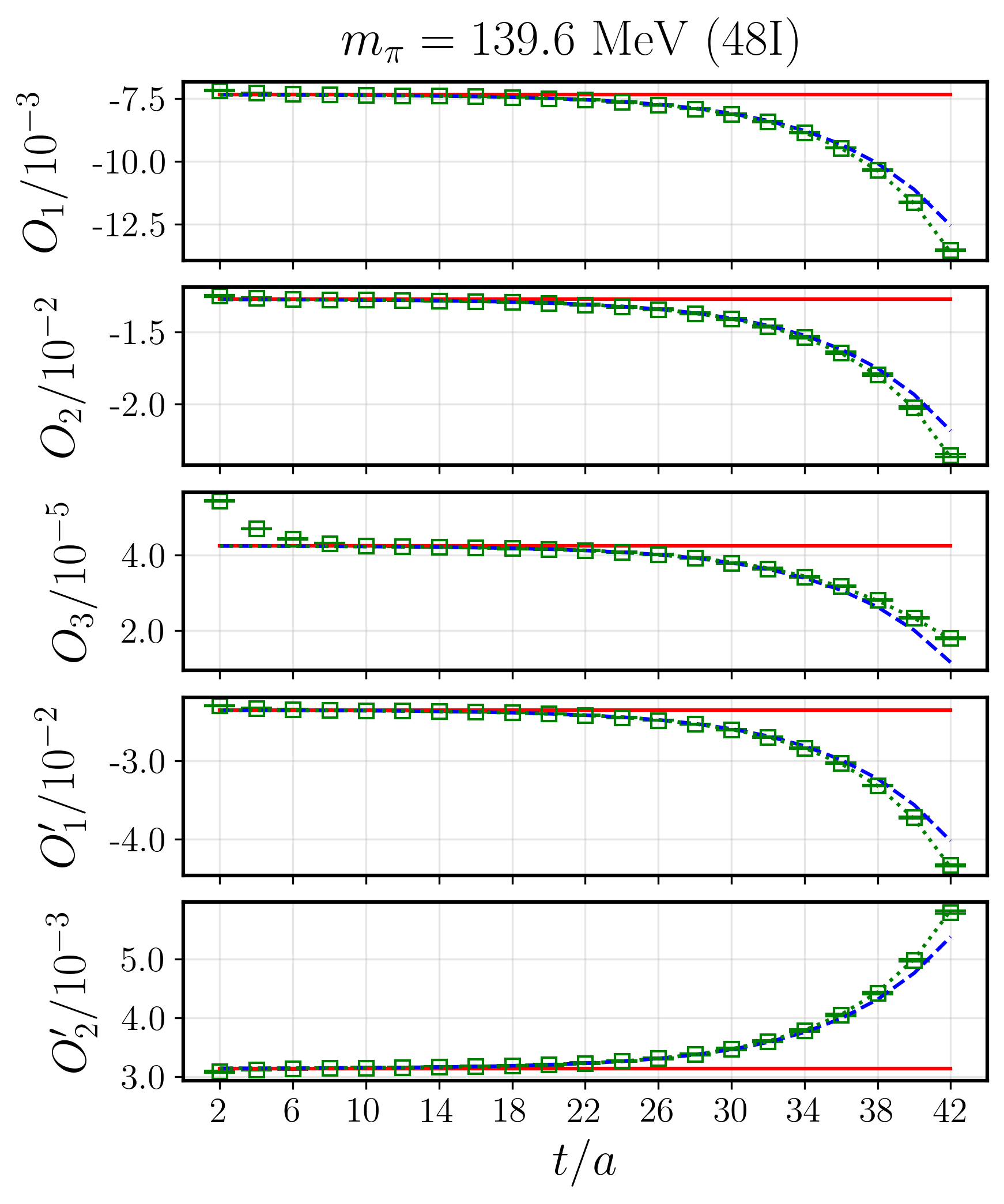}
	\caption{Comparison between the effective bare matrix elements before subtraction of around-the-world effects, \(O_i^{(R_1)}(t)\) (green points) and contributions from diagrams A,B,C, and D in Fig.~\ref{fig:ATW}.
		\label{fig:ATWmodel}
	}
\end{figure}

We select all data points satisfying $d_{\min} \geq \Delta T$ to fit the four parameters $\mathcal{N}_{A,i}$, $\mathcal{N}_{B/C,i}$, $\mathcal{N}_{D,i}$, and $E_{\pi\pi}$. Using the 48I ensemble as an example, the final fit results are summarized in Table~\ref{table:fit}. We choose a sufficiently large separation $\Delta T = 1.8~\text{fm}$ to suppress excited-state contamination. In Fig.~\ref{fig:ATWmodel}, we check the consistency between the lattice data and the fit results of these diagrams. The contribution from $A+B+C+D$ is very consistent with lattice data. For $t\ll T/2$, the around-the-world effects are dominated by diagrams B and C. Thus, to get a plateau for $t\ll T/2$ in the subtraction method, we only need to subtract the contributions from B and C.
\begin{table}
	\centering
    	\begin{tabular}{l|ccccc}
		\hline \hline  &$ \mathcal{O}_1 \times 10^3$&$\mathcal{O}_2 \times 10^3$&$ \mathcal{O}_3 \times 10^5$&$ \mathcal{O}^\prime_{1} \times 10^3$&$\mathcal{O}^\prime_{2} \times 10^3$ \\
		\hline $a^4\mathcal{N}_{A,i}$& $-7.346(19)$ & $-12.743(58)$ & $4.248(15)$ & $-23.493(59)$ & $3.137(15)$  \\
		$a^4\mathcal{N}_{B/C,i}$ & $-6.865(18)$ & $-12.008(54)$ & $-4.075(15)$ & $-22.097(58)$ & $2.961(14)$ \\
		$a^4\mathcal{N}_{D,i}$ & $-7.330(19)$ & $-12.710(58)$ & $4.259(15)$ & $-23.441(59)$  & $3.129(15)$ \\
        \hline
		$aE_{\pi\pi}$ & $0.16141(23)$ & $0.16143(23)$ & $0.16130(23)$ & $0.16141(23)$ & $0.16143(23)$ \\
		\hline \hline
	\end{tabular}
	\caption{\label{table:fit} Direct fit results for the matrix element coefficients corresponding to diagrams A, B, C, and D in Fig.~\ref{fig:ATW} on 48I, under the condition $d_{\min}\geq \Delta T$. We take $a\Delta T=16$, corresponding to $\Delta T=1.8~\text{fm}$. The fit assumes $\mathcal{N}_B = \mathcal{N}_C$.}
\end{table}

From the fit results, we observe that the assumptions $\mathcal{N}_{A,i} \approx \mathcal{N}_{D,i}$ and $aE_{\pi\pi} \approx 2am_\pi = 0.16135(23)$ hold well. For all matrix elements except $\mathcal{O}_3$, we find $\mathcal{N}_{B/C,i} \approx \mathcal{N}_{A,i}$, whereas for $\mathcal{O}_3$, the relation $\mathcal{N}_{B/C,i} \approx -\mathcal{N}_{A,i}$ is satisfied. This explains why the $R_2$ method effectively suppresses the around-the-world effects for all matrix elements except $\mathcal{O}_3$. The coefficient $\mathcal{N}_{A,i}$ is the desired matrix element $\langle \pi|\mathcal{O}_i|\pi\rangle$. The direct fit results for $\mathcal{N}_{A,i}$ agree remarkably well with those obtained from the subtraction method using Eq.~\eqref{Osub2} (see Table~\ref{table:bare_results}), validating the subtraction approach. Since the subtraction method is much simpler, while yielding results consistent with the fit, we adopt it in the main text.

\bibliography{ref}
	
\end{document}